\documentclass[sn-mathphys]{sn-jnl} % Using a specific document class for Springer Nature in Math and Physics

\usepackage[T1]{fontenc} % Using a modern font encoding
\usepackage{lmodern} % Modern font to provide better support
\usepackage{graphicx} % For handling graphics
\usepackage{amsmath, amssymb, amsfonts} % Mathematical symbols and fonts
\usepackage{amsthm} % Theorem environments
\usepackage{mathrsfs} % Provides mathematical script fonts
\usepackage{anyfontsize}
\usepackage[title]{appendix} % Appendices handling
\usepackage{xcolor} % Support for colors
\usepackage{textcomp} % Provides extra symbols
\usepackage{manyfoot} % Supports multiple footnotes
\usepackage{booktabs} % Enhances quality of tables
\usepackage{algorithm} % Environment for algorithms
\usepackage{algpseudocode} % Algorithmic pseudocode
\usepackage{mathtools} % Enhances amsmath by adding new commands
\usepackage{float} % Provides additional options for floating environments
\usepackage{bm} % Provides \bm command for bold mathematics

\usepackage{threeparttable} % Tables with notes
\usepackage{subcaption} % Better handling of subfigures and subtables than subfig
\usepackage[section]{placeins} % Controls float positions within sections
\usepackage{array} % Additional features for arrays and tables
\usepackage{soul} % For text highlighting, underlining, etc.
\begin{document}
\title[Article Title]{A Gaussian Process Based Method with Deep Kernel Learning for Pricing High-dimensional American Options}

\author[1]{\fnm{Jirong} \sur{Zhuang}}\email{yc27478@connect.um.edu.mo}

\author[1]{\fnm{Deng} \sur{Ding}}\email{dding@um.edu.mo}

\author[1]{\fnm{Weiguo} \sur{Lu}}\email{yc07476@connect.um.edu.mo}

\author[1]{\fnm{Xuan} \sur{Wu}}\email{yc27937@connect.um.edu.mo}

\author*[2,3]{\fnm{Gangnan} \sur{Yuan}}\email{gnyuan@gbu.edu.cn}

\affil[1]{\orgdiv{Department of Mathematics}, \orgname{University of Macau}, \orgaddress{ \city{Macau}, \postcode{999078}, \country{China}}}
\affil[2]{\orgname{Great Bay Institute for Advanced Study}, \orgaddress{ \city{Dongguan}, \postcode{523000}, \state{Guangdong}, \country{China}}}

\affil[3]{\orgdiv{School of Mathematics}, \orgname{University of Science and Technology of China}, \orgaddress{ \city{Hefei}, \postcode{230026}, \state{Anhui}, \country{China}}}

\abstract{In this work, we present a novel machine learning approach for pricing high-dimensional American options based on the modified Gaussian process regression (GPR). We incorporate deep kernel learning and sparse variational Gaussian processes to address the challenges traditionally associated with GPR. These challenges include its diminished reliability in high-dimensional scenarios and the excessive computational costs associated with processing extensive numbers of simulated paths Our findings indicate that the proposed method surpasses the performance of the least squares Monte Carlo method in high-dimensional scenarios, particularly when the underlying assets are modeled by Merton's jump diffusion model. Moreover, our approach does not exhibit a significant increase in computational time as the number of dimensions grows. Consequently, this method emerges as a potential tool for alleviating the challenges posed by the curse of dimensionality.}


\maketitle

\section{Introduction}\label{1}
Since the introduction of the Black-Scholes formula in the 1970s, the field of option pricing has advanced significantly. Various traditional numerical methods, including finite difference and tree methods, have been developed to tackle option pricing problems. Among these, pricing American options has emerged as a key area of focus. American options are popular because they can be exercised at any time before their expiration. However, pricing these options, particularly in high-dimensional settings, presents considerable challenges. The computational demand of traditional methods like the tree method grows exponentially with the increase in dimensions, a phenomenon referred to as the curse of dimensionality.\\
\\
A common approach to mitigate the curse of dimensionality is the regression-based Monte Carlo method, which involves simulating numerous paths and then estimating the continuation value through cross-sectional regression to obtain optimal stopping rules. \cite{refCarriere1996} first used spline regression to estimate the continuation value of an option. Inspired by his work, \cite{refTsitsiklis2001} and \cite{refLongstaff2001} further developed this idea by employing least-squares regression. Presently, the Least Squares Method (LSM) proposed by Longstaff and Schwartz has become one of the most successful methods for pricing American options and is widely used in the industry. In recent years, machine learning methods have been considered as potential alternative approaches for estimating the continuation value. Examples include kernel ridge regression \citep{refHan2009,refHu2020}, support vector regression \citep{refLin2021}, neural networks \citep{refKohler2010,refLapeyre2021}, regression trees \citep{refEch2023}, and Gaussian process regression \citep{refLudkovski2018,refGoudenège2019, refGoudenège2020}. In subsequent content, we refer to algorithms that share the same framework as LSM but may utilize different regression methods as Longstaff-Schwartz algorithms. Besides estimating the continuation value, machine learning has also been employed to directly estimate the optimal stopping time \citep{refBecker2019} and to solve high-dimensional free boundary PDEs for pricing American options \citep{refSirignano2018}.
\\
\\
In this work, we will apply a deep learning approach based on Gaussian process regression (GPR) to the high-dimensional American option pricing problem. The GPR is a non-parametric Bayesian machine learning method that provides a flexible solution to regression problems. Previous studies have applied GPR to directly learn the derivatives pricing function \citep{refSpiegeleer2018} and subsequently compute the Greeks analytically \citep{refCrépey2020, refludkovski2021}. This paper focuses on the adoption of GPR to estimate the continuation value of American options. \cite{refLudkovski2018} initially integrated GPR with the regression-based Monte Carlo methods, and testing its efficacy on Bermudan options across up to five dimensions. \cite{refGoudenège2019} further explored the performance of GPR in high-dimensional scenarios through numerous numerical experiments. They also introduced a modified method, the GPR Monte Carlo Control Variate method, which employs the European option price as the control variate. Their method adopts GPR and a one-step Monte Carlo simulation at each time step to estimate the continuation value for a predetermined set of stock prices. In contrast, our study applies a Gaussian-based method within the Longstaff-Schwartz framework, requiring only a global set of paths and potentially reducing simulation costs. Nonetheless, direct integration of GPR with the Longstaff-Schwartz algorithm presents several challenges. First, GPR’s computational cost is substantial when dealing with large training sets, which are generally necessary to achieve a reliable approximation of the continuation value in high dimensional cases. Second, GPR may struggle to accurately estimate the continuation value in high-dimensional scenarios, and we will present a numerical experiment to illustrate this phenomenon in Section \ref{sec5}. 
\\
\\To address the above two challenges, we propose a method for pricing high-dimensional American options that combines the Longstaff-Schwartz algorithm with Deep Kernel Learning (DKL) proposed by \cite{refWilson2016-1,refWilson2016-2}, which is a hybrid of Gaussian process and deep neural networks. Additionally, variational inference \citep{refTitsias2009,refHensman2013} is applied to reduce the computational cost. The paper is organized as follows: Section \ref{sec2} provides a brief introduction to Gaussian process regression. In Section \ref{sec3}, the problem formulation and the Longstaff-Schwartz algorithm are described. Section \ref{section4} introduce the proposed method. Section \ref{sec5} presents the numerical experiments, including the American options under Merton’s jump diffusion model, which has not been discussed in reference \cite{refLudkovski2018,refGoudenège2019}.

\section{Preliminaries on Gaussian Process Regression}\label{sec2}
In this section, we review the basic concepts of Gaussian process regression. Further details can be found in reference \cite{refWilliams2006}.
Consider the following model:
\begin{equation}
    y=f(\mathbf{x})+\epsilon
    \label{eq1}
\end{equation}
where $\mathbf{x} \in\mathbb{R}^{d},y\in\mathbb{R},\epsilon\sim \mathcal{N}(0,\sigma^{2})$ and $f(x)$ is an unknown function. We place a Gaussian process prior on $f$ which is characterized by a mean function $m:\mathbb{R}^{d} \to \mathbb{R}$ and a kernel function $k:\mathbb{R}^{d}\times\mathbb{R}^{d}\to\mathbb{R}$. In this paper, we consider only the Gaussian process prior with zero mean, denoted as $f\sim\mathcal{GP}(\mathbf{0},k)$. Before we proceed to further discussion of the kernel function, let us first establish our regression task.
\\
\\
Consider a training set $\mathcal{D}=(\mathbf{X,y})$ where the components in $\mathbf{y}=[y_{1},\dots,y_{N}]^{\intercal}$ are noisy observations at $N$ locations $\mathbf{X}=[\mathbf{x}_{1},\dots,\mathbf{x}_{N}]^{\intercal}$. We aim to predict the corresponding function values $\mathbf{\tilde f}=[f(\mathbf{\tilde x}_{1}),\dots,f(\mathbf{\tilde x}_{P})]^{\intercal}$ at $P$ unobserved points $\mathbf{\tilde X}=\left[ \mathbf{\tilde x}_{1},\dots,\mathbf{\tilde x}_{P}\right]^\intercal$.
Upon fitting the data to our model (\ref{eq1}), we obtain:
\begin{equation}
    \begin{bmatrix} \mathbf{y}\\ \mathbf{\tilde f}\end{bmatrix} \sim \mathcal{N}
    \left[
    \begin{bmatrix}
        \mathbf{0}_{N}
        \\
        \mathbf{0}_{P}
    \end{bmatrix}
    , \begin{bmatrix} 
            K(\mathbf{X,X})+\sigma \mathbf{I} & K(\mathbf{X,\tilde X}) \\
            K(\mathbf{\tilde X,X}) & K(\mathbf{\tilde X,\tilde X})
        \end{bmatrix}
    \right]
\end{equation}
where $K(\mathbf{X,X})$ denotes the covariance matrix with entries $K(\mathbf{X,X})_{i,j}=k(\mathbf{x}_{i},\mathbf{x}_{j})$, and $I$ is an identity matrix. Consequently, the conditional posterior $\mathbf{\tilde f}|\mathbf{X,y},\mathbf{\tilde X} \sim \mathcal{N}(\mu(\mathbf{\tilde X}) ,\Sigma(\mathbf{\tilde X}))$, where

\begin{equation}\label{eq3}
    \mu(\mathbf{\tilde X}) \coloneqq \mathbb{E} \left[\mathbf{\tilde f}|\mathbf{X,y},\mathbf{\tilde X} \right] = K(\mathbf{\tilde X,X})[ K(\mathbf{X,X})+\sigma \mathbf{I} ]^{-1}\mathbf{y}
\end{equation}
\begin{equation}
    \Sigma(\mathbf{\tilde X}) \coloneqq {\rm cov}\left[\mathbf{\tilde f}|\mathbf{X,y},\mathbf{\tilde X} \right]= K(\mathbf{\tilde X,\tilde X})- K(\mathbf{\tilde X,X})[ K(\mathbf{X,X})+\sigma \mathbf{I} ]^{-1}K(\mathbf{X,\tilde X}).
\end{equation}
Or equivalently, the posterior mean function can be expressed as a linear combination of kernel functions:
\begin{equation}
    \mu(\mathbf{\tilde x})=\sum^{N}_{i=1}\alpha_{i}k(\mathbf{x}_{i},\mathbf{\tilde x}) \quad \text{where} \quad [\alpha_1,\dots,\alpha_N]^\intercal=(K(\mathbf{X,X})+\sigma^{2}\mathbf{I})^{-1}\mathbf{y}.
\end{equation} 
Returning to the discussion of the kernel function, given any compact subset $U\subset\mathbb{R}^{d}$, and denote the space of all continuous functions on $U$ by $\mathbf{C}(U)$. Then, for certain kernels $k$, the space generated by the kernel functions ${\rm span}\left\{ k(\mathbf{x},\cdot): \mathbf{x} \in U \right\}$ is dense in $\mathbf{C}(U)$ in maximum norm. This kind of kernels is called the universal kernel \citep{refMicchelli2006}, and an important example is the RBF kernel:
\begin{equation}
    k_{{\rm RBF}}\left(\mathbf{x,x^\prime}\right)=\exp\left(-\frac{||\mathbf{x-x}^{\prime}||^2}{\gamma^{2}}\right)\quad \text{where} \;\;  \mathbf{x,x}^\prime\in \mathbb{R}^{d}, \;\; ||\cdot|| \, \text{is Euclidean norm}
\end{equation}
\\
The parameter $\gamma$ is known as the length-scale parameter, which controls the rate at which the kernel function’s value changes with the input. In this model, several (hyper-)parameters need to be determined, including the standard deviation of the noise $\sigma$ and the length-scale parameter $\gamma$ of the RBF kernel. These (hyper-)parameters can be selected by setting the log-likelihood function
\begin{equation}\label{eq7}
    \log \mathbb{P}(\mathbf{y|X})=-\frac{1}{2}\log\left[\det(K(\mathbf{X,X})+\sigma^{2} \mathbf{I})\right]-\frac{1}{2}\mathbf{y}^{\intercal}[K(\mathbf{X,X})+\sigma^{2}\mathbf{I}]^{-1}\mathbf{y}-\frac{N}{2}\log(2\pi)
\end{equation}as the objective function and then maximize it using various gradient descent methods. Note that the computational complexity of GPR is $O(N^{3})$, due to the inverse of a $N \times N$ matrix are included in equation (\ref{eq3}) and equation (\ref{eq7}).

\section{Problem Formulation and Longstaff-Schwartz Algorithm}\label{sec3}

Let $(\Omega,\mathcal{F},\left\{\mathcal{F}_{t}\right\}_{t\ge 0},\mathbb{Q})$ be a filtered probability space with risk-neutral measure. Consider an American option on $d$ underlying assets $( \mathbf{S}_{t} )_{t \ge 0} \subset \mathbb{R}^{d}$ with expiration date $T$ and payoff function $h(\cdot)$, and we assume the prices of the underlying assets follow a Markovian process. Let the constant $r$ denote the risk-free interest rate and define the discount factor $D_{s,t}\coloneqq e^{(s-t)r}$. Then, the option value at time $t$ is $V_{t}(\mathbf{S}_{t})=\sup_{\tau\in\mathcal{T}, \tau>t}\mathbb{E}[D_{t,\tau}h(\mathbf{S}_{\tau})]$ where $\mathcal{T}$ denotes all stopping times with respect to $\mathcal{F}$. In particular, we are interested in the price of the option at time $0$ which is the solution of an optimal stopping problem $V_{0}=\sup_{\tau \in \mathcal{T}}\mathbb{E}[ D_{0,\tau}h_{\tau}(\mathbf{S}_{\tau})]$. The main idea of solving it is to determine the optimal early exercise policy by using backward dynamic programming recursion. In practice, we will consider a Bermudan option with exercise dates $t_{1}<t_{2} < \dots < t_{n}=T$ to estimate the price of the American option. For simplification of the notation, we use $\mathbf{S}_i$ as the shorthand for the underlying assets price at $t_i$ and $D_{i,i+1}\coloneqq e^{(t_i-t_{i+1})r}$ denote the discount factor in the following contents. Let $\mathbb{T}=\{ 1,\dots,n \}$. Then, we consider the following backward dynamic programming recursion:
$$\left \{ \begin{array}{lcl}
      V_{n}(\mathbf{S}_{n})=h(\mathbf{S}_{n})
     & \\V_{i}(\mathbf{S}_{i})=\max \{h(\mathbf{S}_{i}),Q_{i}(\mathbf{S}_{i})\} \text{ for } i=n-1,n-2,\dots,1
\end{array} \right.
$$
where $Q_{i}(\mathbf{S}_{i}) \coloneqq D_{i,\,i+1}\mathbb{E}[V_{i+1}| \mathbf{S}_{i}]$ represents the discounted continuation value of the option at $t_{i}$.
\\
\\Equivalently, find the optimal exercise date $\hat{\tau}_1=\min \left\{\tau \in \mathbb{T} \colon h(\mathbf{S}_{\tau})\ge Q_{\tau}(\mathbf{S}_{\tau})   \right\}$ via:
$$\left \{ \begin{array}{lcl}
      \hat{\tau}_{n}=n
     & \\ \hat{\tau}_{i}=i \cdot 1_{\{h(\mathbf{S}_{i}) > Q_{i}(\mathbf{S}_{i})\} }+\hat{\tau}_{i+1} 1_{\{h(\mathbf{S}_{i}) \leq Q_{i}(\mathbf{S}_{i})\}} \text{ for } i=n-1,n-2,\dots,1.
\end{array} \right.
$$
\\A key task in this backward Dynamic Programming recursion is to find $Q_{i}(\mathbf{S}_{i})$. \cite{refLongstaff2001} proposed the least square Monte Carlo method which estimates the continuation value of the option by a least-square regression. Firstly, simulate $N$ independent paths $\{ \mathbf{S}^{(j)}_{0}, \mathbf{S}^{(j)}_{1},\dots,\mathbf{S}^{(j)}_{n}\}$ for $j=1,\dots,N$, and then estimate the optimal exercise date by the following policy:
$$\left \{ \begin{array}{lcl}
      \hat{\tau}^{(j)}_{n}=n
     & \\ \hat{\tau}^{(j)}_{i}=i \cdot 1_{\{h(\mathbf{S}^{(j)}_{i}) > \hat{Q}_{i}(\mathbf{S}^{(j)}_{i})\} }+\hat{\tau}^{(j)}_{i+1} 1_{\{h(\mathbf{S}^{(j)}_{i}) \leq \hat{Q}_{i}(\mathbf{S}^{(j)}_{i})\}} \text{ for } i=n-1,n-2,\dots,1.
\end{array}  \right.$$
\\For each step, we set $$\mathbf{X}_i=\left[\vec{\phi}(\mathbf{S}_{i}^{(1)}),\dots,\vec{\phi}(\mathbf{S}_{i}^{(N)})\right]^{\intercal}$$ and $$\mathbf{y}_i=\left[V_{i+1}(\mathbf{S}_{i+1}^{(1)}),\dots,V_{i+1}(\mathbf{S}_{i+1}^{(N)})\right]^{\intercal}$$
where $V_{i+1}(\mathbf{S}_{i+1}^{(j)})\coloneqq D_{i+1,\,\hat{\tau}_{i+1}} h(\mathbf{S}_{\hat{\tau}_{i+1}}^{(j)})$, 
$\vec{\phi}(\mathbf x)=[\phi_{1}(\mathbf x),\dots,\phi_{J}(\mathbf x)]^{\intercal}$\hspace{0.3cm} $(\mathbf x\in \mathbb{R}^{d}, J \in \mathbb{N}^{+})$ and $\phi_{1}(\cdot),\dots,\phi_{J}(\cdot)$ are basis functions, e.g., polynomials, Laguerre polynomials etc. Next, solve the following least-squares problem
\begin{equation}\label{eq8}
    \hat{f}_i=\mathop{\rm arg\;min}\limits_{f\in {\rm span}\{\phi_{1},\dots,\phi_{J}\}} \frac{1}{N} \sum^{N}_{j=1}\left(f(\mathbf{S}^{(j)}_{i})-V_{i+1}(\mathbf{S}_{i+1}^{(j)})\right)^{2} .
\end{equation}
\\Then the estimated discounted continuation value is $\hat{Q}_{i}(\mathbf{S}_{i})\coloneqq D_{i,\,i+1} \hat{f}_i(\mathbf{S}_{i})=D_{i,\,i+1}\vec{\phi}(\mathbf{S}_{i})^{\intercal}\mathbf{b}_i$, where $\mathbf{b}_i=(\mathbf{X}_i^{\intercal}\mathbf{X}_i)^{-1}\mathbf{X}_i^{\intercal}\mathbf{y}_i$. Finally, the option price at time $0$ is estimated by  $\hat{V}_{0}=\frac{D_{0,1}}{N}\sum^{N}_{j=1}V_{1}(\mathbf{S}_{1}^{(j)})$.  Note that we need to calculate the inverse of a $J\times J$ matrix at each time step of the backward dynamic programming recursion, and the computational complexity of this task is $O(J^{3})$, where $J$ can be very large in pricing high-dimensional options. Additionally, the choice of basis functions is another crucial issue. It can significantly affect the accuracy of the pricing, yet there is no objective method for selecting them. Therefore, Gaussian process based models are considered in estimating the continuation value.

\section{Longstaff-Schwartz Algorithm with Deep Kernel Learning}\label{section4}

Inspired by a version of Regression Monte Carlo Methods, proposed by \cite{refLudkovski2018}, which applied GPR to estimate the continuation value within the Longstaff-Schwartz algorithm in low-dimensional cases, we aim to extend the application of GPR to the pricing of high-dimensional American options. Prior to this, there are two potential challenges in high-dimensional scenarios. The first challenge is the unreliability of the Euclidean metric in high-dimensional spaces. Common kernels, such as the Radial Basis Function (RBF) kernel, employ the Euclidean metric to measure the similarity between two inputs; however, this may not be optimal for high-dimensional data \citep{refAggarwal2001}.\\
\\~
To overcome this issue, we incorporate the Deep Kernel Learning (DKL) method, introduced by \cite{refWilson2016-1,refWilson2016-2}, which employs a deep neural network to learn a non-Euclidean metric for the kernel. The second challenge is the significant computational cost associated with processing numerous simulated paths. A large training set is generally required to achieve a reliable approximation in high-dimensional option pricing, and the computational complexity of GPR is $O(N^{3})$, where $N$ is the size of the training set. Given the substantial computational expense, we consider the adoption of a sparse variational Gaussian process. For further details on sparse variational Gaussian processes, we refer to \cite{refTitsias2009,refHensman2013,refhensman2015}. In the subsequent two subsections, we will demonstrate how these machine learning techniques can be integrated into the Longstaff-Schwartz algorithm.

\subsection{DKL Model for Estimating Continuation Values }\label{subsection3.3.1}
\begin{figure}[htb]
    \centering
    \includegraphics[width=1\textwidth]{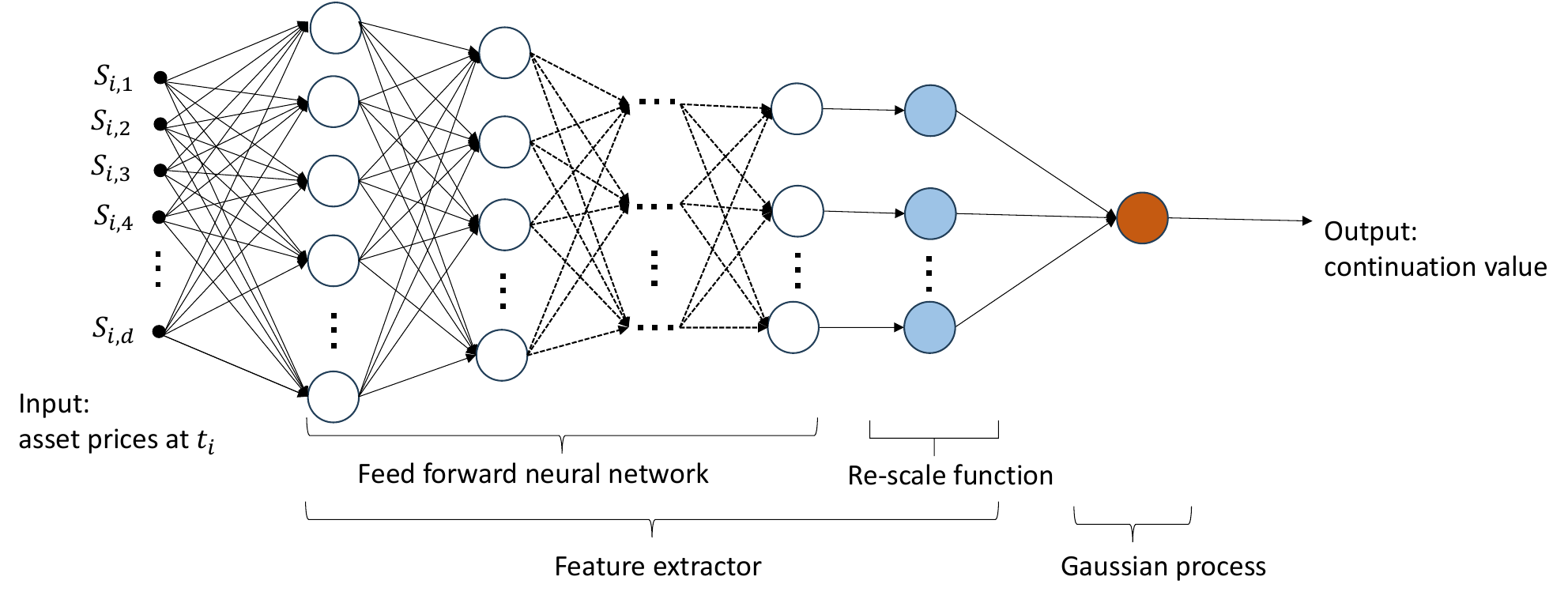}
    \caption{Structure of the Deep Kernel Learning Model}
    \label{fig_Structure}
\end{figure}

In this subsection, we present an overview of the structure of the DKL model that we will use for estimating continuation values. Consider the following DKL model $\Gamma$:
\begin{equation}
    y=\Gamma(\mathbf x)+\epsilon \coloneqq f \circ g^{\theta}(\mathbf x)+\epsilon
\end{equation}
where $\mathbf x \in\mathbb{R}^{d}$ is the vector of prices for $d$-underlying assets, $y\in\mathbb{R}$ is the continuation value of the option, $\epsilon\sim \mathcal{N}(0,\sigma^{2})$ is the noise, $g^{\theta}\colon\mathbb{R}^{d} \xrightarrow{} \mathbb{R}^{d^{\prime}} \quad (d^\prime < d)$ is a feature extractor that contains a feedforward neural network and a re-scale function (see Figure \ref{fig_Structure}). $f\colon\mathbb{R}^{d^{\prime}} \xrightarrow{} \mathbb{R}$ is an unknown function with Gaussian process prior $f\sim\mathcal{GP}(\mathbf{0},k^{\theta_{\rm ker}})$. $\theta_{\rm ker}$ and $\theta$ represent the kernel parameters and neural network parameters, respectively. Note that we may simply use the notation $k$ to represent $k^{\theta_{\rm ker}}$ to simplify the notation in this section.
\\
\\Let's discuss the feature extractor first.
$g^{\theta}\coloneqq l \circ \Psi^{\theta}$ where $\Psi^{\theta}:\mathbb{R}^{d} \xrightarrow{} \mathbb{R}^{d^{\prime}}$ is a $L$-layers fully connected neural network of the form 
$$\Psi^{\theta}\coloneqq a_{L}\circ \psi_{L-1} \circ a_{L-1} \circ \dots  \psi_{1} \circ a_{1} $$ 
\begin{itemize}
    \item $a_{1} :\mathbb{R}^{d} \xrightarrow{} \mathbb{R}^{d_{1}}$, $a_{2}                 :\mathbb{R}^{d_{1}} \xrightarrow{} \mathbb{R}^{d_{2}}$,$\dots$, $a_{L} :\mathbb{R}^{d_{L-1}}          \xrightarrow{} \mathbb{R}^{d^{\prime}}$ are affine functions
    \\$a_j(\mathbf{x})=W_j\mathbf{x}+\bm{\beta}_j$ where $W_j\in\mathbb{R}^{d_j\times d_{j-1}}$, $\bm{\beta}_j\in\mathbb{R}^{d_j}$ are weighted matrices and bias vectors respectively
    \item $\theta=\left( W_j, \bm{\beta}_j\right)^{L}_{j=1}$ are the parameters in the neural network
    \item  $\psi_{1} :\mathbb{R}^{d_{1}} \xrightarrow{} \mathbb{R}^{d_{1}}$, $\psi_{2} :\mathbb{R}^{d_{2}} \xrightarrow{} \mathbb{R}^{d_{2}}$, $\dots$,$\psi_{L-1} :\mathbb{R}^{d_{L-1}} \xrightarrow{} \mathbb{R}^{d_{L-1}}$ are ReLU activation functions  ($\psi_j(\mathbf{x})=[x^+_1,x^+_2,\dots,x^+_{d_j}]^\intercal$ where $x_1,\cdots,x_{d_j}$ are components of $\mathbf{x}$).
\end{itemize}
And $l(\cdot):\mathbb{R}^{d^{\prime}} \xrightarrow{} \mathbb{R}^{d^{\prime}}$ is a function that re-scale the output of $\Psi^{\theta}$ into the range $[-1,1]^{d^{\prime}}$. The $\eta$-th element of the vector $l(\Psi^{\theta}(\mathbf x))\in \mathbb{R}^{d^{\prime}}$ is

\begin{equation*}
[l(\Psi^{\theta}(x))]_{\eta}=
\begin{cases}

-1, & [\Psi^{\theta}(\mathbf x)]_{\eta} < \mathop{\min}\limits_{j=1,\dots,N}[\Psi^{\theta}(\mathbf{x}_{j})]_{\eta},\\
1, & [\Psi^{\theta}(\mathbf x)]_{\eta} > \mathop{\max}\limits_{j=1,\dots,N}[\Psi^{\theta}(\mathbf{x}_{j})]_{\eta},\\
2 \cdot \frac{[\Psi^{\theta}(\mathbf x)]_{\eta}-\mathop{\min}\limits_{j=1,\dots,N}[\Psi^{\theta}(\mathbf{x}_{j})]_{\eta}}{\mathop{\max}\limits_{j=1,\dots,N}[\Psi^{\theta}(\mathbf{x}_{j})]_{\eta}-\mathop{\min}\limits_{j=1,\dots,N}[\Psi^{\theta}(\mathbf{x}_{j})]_{\eta}}-1, & \text{otherwise}.
\end{cases}
\end{equation*}
\\
\\The purpose of the feature extractor is to project the inputs from a high-dimensional space to a low-dimensional latent space. Then, we can apply the Gaussian process with the conventional RBF kernel in this latent space. The whole model can also be interpreted as a Gaussian process with the deep kernel. For example, if we adopt the RBF kernel, i.e. $k^{\theta_{\rm ker}}(\cdot,\cdot)=k^{\theta_{\rm ker}}_{\rm RBF}(\cdot,\cdot)$, then
\begin{equation*}
    k^{\theta_{\rm ker}}_{\rm deep}(\mathbf {x,x}^{\prime})=k^{\theta_{\rm ker}}_{\rm RBF}\left(g^{\theta}(\mathbf x),g^{\theta}(\mathbf{x}^{\prime})\right)=\exp\left(-\frac{||g^{\theta}(\mathbf x)-g^{\theta}(\mathbf{x}^{\prime})||^2}{\gamma^{2}}\right)\quad \mathbf{x,x}^\prime\in \mathbb{R}^{d}
\end{equation*} 
As a result, the similarity of two data points $\mathbf{x,x^{\prime}}$ is measured by a non-Euclidean metric $||\mathbf{x}-\mathbf{x}^{\prime}||_{\rm deep} \coloneqq ||g^{\theta}(\mathbf x)-g^{\theta}(\mathbf{x}^{\prime})||$ instead of the Euclidean metric in a conventional RBF kernel. The purpose of training the neural network $g^{\theta}(\cdot)$ can be seen as learning a suitable non-Euclidean metric for the kernel.

\subsection{Proposed Method}
\begin{figure}[htb]
    \centering
    \includegraphics[width=1\textwidth]{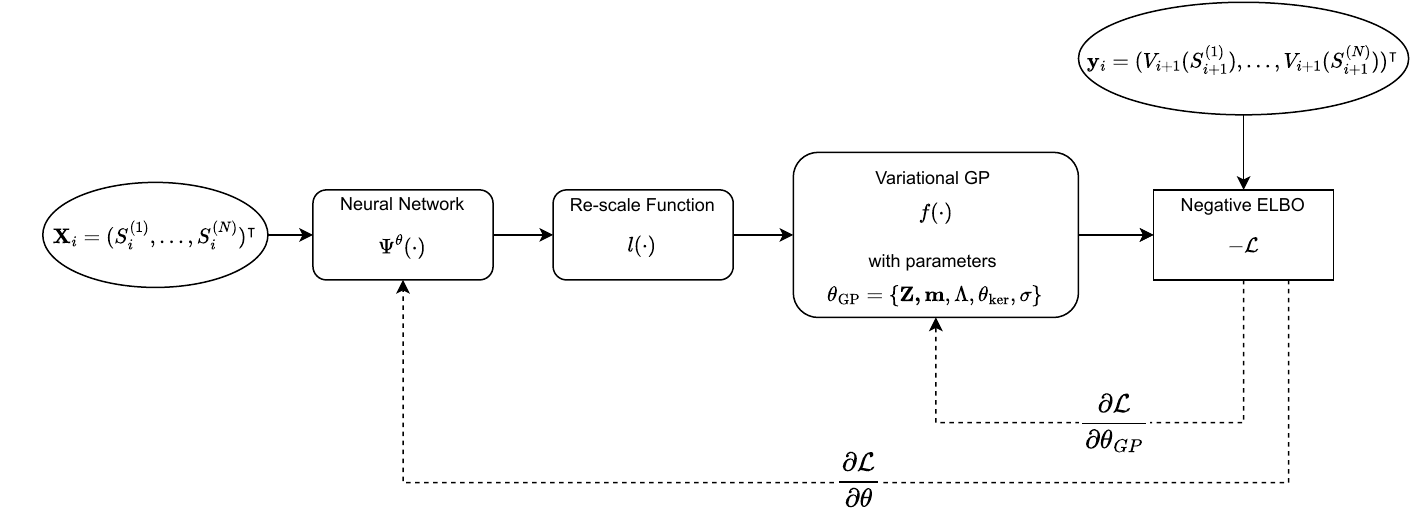}
    \caption{Updating the parameters of DKL model}
    \label{fig_flow}
\end{figure}

In this subsection, we discuss the implementation of the sparse variational Gaussian process. Then, we show how the model is trained and provide the continuation value needed at each step. Finally, the full algorithm is provided in pseudocode.
\\
\\At time step $t_{i}$, we adopt $(\mathbf{X}_i,\mathbf{y}_i)$ as the training set of $\Gamma$ where $$\mathbf{X}_i=\left[\mathbf{x}_{1},\dots,\mathbf{x}_{N}\right]^{\intercal}\coloneqq\left[\mathbf{S}_{i}^{(1)},\dots,\mathbf{S}_{i}^{(N)}\right]^{\intercal},$$  $$\mathbf{y}_i=\left[y_{1},\dots,y_{N}\right]^{\intercal}\coloneqq\left[V_{i+1}(\mathbf{S}_{i+1}^{(1)}),\dots,V_{i+1}(\mathbf{S}_{i+1}^{(N)})\right]^{\intercal}.$$ 
Denote $\mathbf{G}^{\theta}_i=\left[g^{\theta}(\mathbf{x}_{1}),\dots,g^{\theta}(\mathbf{x}_{N})\right]^{\intercal}$ as the latent vectors of the inputs $\mathbf{X}_i$, and $\mathbf{f}^{\theta}_i=\left[f\circ g^{\theta}(\mathbf{x}_{1}),\dots,f\circ g^{\theta}(\mathbf{x}_{N})\right]^{\intercal}$ as the corresponding function values.
\\
\\For the Gaussian process in the DKL model, we adopt the sparse variational Gaussian process. The idea of the sparse variational Gaussian process is that, a variational posterior is constructed to approximate the true posterior, reducing the computational complexity from $O(N^3)$ to $O(NM^2)$, where $M$ is the number of the inducing points.  We now define a set of $M$ inducing points $\left \{ \mathbf{z}_{1},\dots,\mathbf{z}_{M} \right \} \subset\mathbb{R}^{d^{\prime}}$ $(M \ll N)$ for constructing our variational posterior. Note that the function $l(\cdot)$ ensures that points $g^{\theta}(\mathbf{x}_{1}),\dots,g^{\theta}(\mathbf{x}_{N})$ lie within the hyper-cube $[-1,1]^{d^{\prime}}\subset \mathbb{R}^{d^\prime}$. Hence, we can initialize the locations of the inducing points randomly in or around that cube; here, we initialize them with $\mathbf{z}_j\sim\mathcal{N}(\mathbf{0},\mathbf{I}_{d^\prime})$. Let $\mathbf{Z}=\left[\mathbf{z}_{1},\dots,\mathbf{z}_{M}\right]^{\intercal}$ and its corresponding function values $\mathbf{u}=\left[f(\mathbf{z}_{1}),\dots,f(\mathbf{z}_{M})\right]^{\intercal}$. The locations of these inducing points are parameters that need to be optimized later. 
\\
\\
Suppose we would like to make a prediction on a point $\mathbf{\tilde x}\in\mathbb{R}^d$. We define a variational posterior $q(\Gamma(\mathbf{\tilde x}),\mathbf{u})=p(\Gamma(\mathbf{\tilde x})|\mathbf{u})q(\mathbf{u})$. For $q(\mathbf{u})$, we follow the strategy proposed by \cite{refHensman2013} that lets $q(\mathbf{u})=\mathcal{N}(\mathbf{u}|\mathbf{m},\Lambda)$. The mean vector $\mathbf{m}$ and covariance matrix $\Lambda$ are parameters that need to be learned later. For $p(\Gamma(\mathbf{\tilde x})|\mathbf{u})$, we consider the joint distribution of $\mathbf{u}$ and $\Gamma(\mathbf{\tilde x})$:
\begin{equation}
    \begin{bmatrix} \mathbf{u}\\ \Gamma(\mathbf{\tilde x})\end{bmatrix} \sim \mathcal{N}
    \left[
    \begin{bmatrix}
        \mathbf{0}_M
        \\
        \mathbf{0}
    \end{bmatrix}
    , \begin{bmatrix} 
            K(\mathbf{Z,Z}) & k(\mathbf{Z}, g^\theta(\mathbf{\tilde x})) \\
            k(g^\theta(\mathbf{\tilde x}),\mathbf{Z}) & k(g^\theta(\mathbf{\tilde x}),g^\theta(\mathbf{\tilde x}))
        \end{bmatrix}
    \right]
\end{equation}
where $k(\mathbf{Z}, g^\theta(\mathbf{\tilde x}))=[k(\mathbf{z}_1,g^\theta(\mathbf{\tilde x})),\dots,k(\mathbf{z}_M, g^\theta(\mathbf{\tilde x}))]^\intercal$.
As a consequence, $p(\Gamma(\mathbf{\tilde x})|\mathbf{u})= \mathcal{N}(\Gamma(\mathbf{\tilde x})|\mu ,\Sigma)$, where
\begin{equation}
    \mu = k(g^\theta(\mathbf{\tilde x}),\mathbf{Z})K(\mathbf{Z,Z})^{-1}\mathbf{u},
\end{equation}
\begin{equation}
     \Sigma =k(g^\theta(\mathbf{\tilde x}),g^\theta(\mathbf{\tilde x}))- k(g^\theta(\mathbf{\tilde x}),\mathbf{Z})K(\mathbf{Z,Z})^{-1} k(\mathbf{Z},g^\theta(\mathbf{\tilde x}))
\end{equation}
\\
Let $\theta_{\rm GP}=\left(\mathbf{Z,m},\Lambda,\theta_{\rm ker},\sigma \right)$ represent the collection of all parameters in the Gaussian process part of the model, and then marginalize $\mathbf{u}$:
\begin{equation}
    q(\Gamma(\mathbf{\tilde x}))=\int p(\Gamma(\mathbf{\tilde x}) |\mathbf{u}) q(\mathbf{u}) \, d\mathbf{u}=\mathcal{N}(\Gamma(\mathbf{\tilde x})|\tilde\mu,\tilde\Sigma)
\end{equation}
where 
\begin{equation}
    \tilde\mu =\tilde\mu_{\theta,\theta_{\rm GP}}(\mathbf{\tilde x}) \coloneqq k(g^{\theta}(\mathbf{\tilde x}),\mathbf{Z})K(\mathbf{Z,Z})^{-1}\mathbf{m},
\end{equation}
\begin{equation}
     \tilde\Sigma=k(g^{\theta}( \mathbf{\tilde x}), g^{\theta}( \mathbf{\tilde x}))- k(g^{\theta}( \mathbf{\tilde x}),\mathbf{Z})K(\mathbf{Z,Z})^{-1} (K(\mathbf{Z,Z})-\Lambda)K(\mathbf{Z,Z})^{-1}k(\mathbf{Z},g^{\theta}(\mathbf{\tilde x})).
\end{equation}
\\
The variational posterior mean can be used to make our predictions. Before that, we need to train the model first.
\\
\\Consider the Kullback-Leibler divergence $\mathbf{KL}[q|| p]$ which measures the difference between the variational posterior and the true posterior:  
\begin{align}
        \mathbf{KL}[q(\mathbf{f}_i^\theta,\mathbf{u})|| p(\mathbf{f}_i^\theta,\mathbf{u}| \mathbf{y}_i)] &= \int \int q(\mathbf{f}_i^\theta,\mathbf{u})\,\log\left(\frac{q(\mathbf{f}_i^\theta,\mathbf{u})}{p(\mathbf{f}_i^\theta,\mathbf{u}|\mathbf{y}_i)}\right) \,d\mathbf{f}^\theta_i\,d\mathbf{u}
        \\&=-\mathcal{L}(\theta,\theta_{\rm GP})+\log p(\mathbf{y}_i)\label{eq_kl}
\end{align}
 where $\mathcal{L}(\theta,\theta_{\rm GP})\coloneqq\int \int q(\mathbf{f}_i^\theta,\mathbf{u})\log\left(\frac{p(\mathbf{f}_i^\theta,\mathbf{u},\mathbf{y}_i)}{q(\mathbf{f}_i^\theta,\mathbf{u})}\right) \,d\mathbf{f}_i^\theta \,d\mathbf{u}$ is called the Evidence Lower Bound (ELBO). From the Equation \ref{eq_kl}, we can deduce that minimizing the negative ELBO $-\mathcal{L}(\theta,\theta_{\rm GP})$ is equivalent to minimizing $\mathbf{KL}[q||p]$. To make it clearer, $\mathcal{L}(\theta,\theta_{\rm GP})$ can be rewritten in the form of
  \begin{align}
     \mathcal{L}(\theta,\theta_{\rm GP})&=\mathbb{E}_{q(\mathbf{f}_i^\theta,\mathbf{u})} \left[\log p(\mathbf{y}_i|\mathbf{f}_i^\theta,\mathbf{u})\right]+\mathbb{E}_{q(\mathbf{f}_i^\theta,\mathbf{u})} \left[\log \frac{p(\mathbf{f}_i^\theta,\mathbf{u})}{q(\mathbf{f}_i^\theta,\mathbf{u})}\right]
     \\&=\mathbb{E}_{p(\mathbf{f}_i^\theta|\mathbf{u})q(\mathbf{u})} \left[\log p(\mathbf{y}_i|\mathbf{f}_i^\theta)\right]+\mathbb{E}_{p(\mathbf{f}_i^\theta|\mathbf{u})q(\mathbf{u})} \left[\log \frac{p(\mathbf{f}_i^\theta|\mathbf{u})p(\mathbf{u})}{p(\mathbf{f}_i^\theta|\mathbf{u})q(\mathbf{u})}\right]
     \\&=\mathbb{E}_{q(\mathbf{f}_i^\theta)}\left[ \log p(\mathbf{y}_i|\mathbf{f}_i^\theta)\right]- \mathbf{KL}\left[q(\mathbf{u})|| p(\mathbf{u})\right]
 \end{align} 
where $q(\mathbf{f}_i^\theta)=\int p(\mathbf{f}_i^\theta|\mathbf{u}) q(\mathbf{u}) \, d\mathbf{u}$. Then, the  variational parameters $\mathbf{Z,m},\Lambda$ can be learned jointly with the neural network parameters $\theta$, standard deviation of the noise $\sigma$ and kernel parameters $\theta_{\rm ker}$ via an optimization problem:
 \begin{equation}
     \theta^\ast,\theta^\ast_{\rm GP}=\mathop{\rm arg\;min}\limits_{\theta,\theta_{\rm GP}} -\mathcal{L}(\theta,\theta_{\rm GP}).
 \end{equation}
In general, this optimization problem doesn't has an analytical solution, therefore, we adopt the backpropagation and gradient based descent optimizer to solve it approximately (see Figure \ref{fig_flow}). Practically, this can be achieved by using state-of-art machine learning packages (e.g. PyTorch \citep{refpytorch2019}). The optimized variational posterior mean function $$\hat{\Gamma}(\mathbf{\tilde x}) \coloneqq \tilde\mu_{\theta^\ast,\theta^\ast_{\rm GP}}(\mathbf{\tilde x})=k(g^{\theta^\ast}(\mathbf{\tilde x}),\mathbf{Z}^\ast)K(\mathbf{Z^\ast,Z^\ast})^{-1}\mathbf{m}^\ast$$ can be used to approximate the continuation value function of the option at time $t_{i}$ where the parameters with asterisks represent the optimized parameters. 
 
Finally, we present the algorithm \ref{alg:1} for using DKL model that we have introduced in previous parts to price American options:
\begin{algorithm}[H]
\caption{Longstaff-Schwartz algorithm with Deep Kernel Learning}
\label{alg:1}
    \begin{algorithmic}
        \State Simulate $N$ independent paths $\left\{\mathbf{\mathbf{S}}^{(j)}_{0}, \mathbf{\mathbf{S}}^{(j)}_{1},\dots,\mathbf{\mathbf{S}}^{(j)}_{n}\right\}$ for $j=1,\dots,N$
        \State Set $V_{n}(\mathbf{\mathbf{S}}^{(j)}_{n})=h(\mathbf{\mathbf{S}}^{(j)}_{n})$ for $j=1,\dots,N$
        \For{$i=n-1,n-2,\dots,1$}
            \State Initialize DKL model $\Gamma$
            \State Set $\mathbf{X}_{i}=\left\{\mathbf{\mathbf{S}}^{(1)}_{i},\mathbf{\mathbf{S}}^{(2)}_{i},\dots,\mathbf{\mathbf{S}}^{(N)}_{i}\right\}$ and \State\hspace{0.6cm}$\mathbf{y}_{i}=\left\{ V_{i+1}(\mathbf{\mathbf{S}}^{(1)}_{i+1}),V_{i+1}(\mathbf{\mathbf{S}}^{(2)}_{i+1}),\dots,V_{i+1}(\mathbf{\mathbf{S}}^{(N)}_{i+1})\right\}$
            \State Train $\Gamma$ on the training set $(\mathbf{X}_{i},\mathbf{y}_{i})$ 
            \For {all paths $\mathbf{\mathbf{S}}^{(j)}_{i}$ in $\left\{ \mathbf{\mathbf{S}}^{(1)}_{i},\dots,\mathbf{\mathbf{S}}^{(N)}_{i}\right\}$}
                \If {$h(\mathbf{\mathbf{S}}^{(j)}_{i})>0$ \bf{and} $h(\mathbf{\mathbf{S}}^{(j)}_{i})>D_{i,\,i+1}\cdot \hat{\Gamma}(\mathbf{\mathbf{S}}^{(j)}_{i})$}
                    \State $V_{i}(\mathbf{\mathbf{S}}^{(j)}_{i})=h(\mathbf{\mathbf{S}}^{(j)}_{i})$
                \Else
                    \State $V_{i}(\mathbf{\mathbf{S}}^{(j)}_{i})=D_{i,\,i+1}\cdot V_{i+1}(\mathbf{\mathbf{S}}^{(j)}_{i+1})$
                \EndIf
            \EndFor

        \EndFor
        \State\Return $V_{0}=\frac{D_{0,1}}{N}\sum^{N}_{j=1}V_{1}(\mathbf{S}^{(j)}_{1})$
    
    \end{algorithmic}
\end{algorithm}

\section{Numerical experiments}\label{sec5}

In this section, we introduce a suite of numerical experiments designed to evaluate the efficacy of Deep Kernel Learning (DKL) for the pricing of American options under geometric Brownian motion (GBM) and Merton's jump diffusion model (MJD). All experiments are performed using Python 3.9.13 on a personal computer with Intel® Core™ i7-12700 CPU, NVIDIA GeForce RTX3060 GPU and 32GB RAM. The deep kernel learning methods are implemented by using machine learning packages PyTorch \citep{refpytorch2019} and Gpytorch \citep{refGardner2018} and are trained using the optimizer provided by torch.optim.sgd in PyTorch, with a learning rate $0.1$, momentum $0.9$, weight decay $10^{-8}$ and a total of $1500$ iterations. Least squares regressions are implemented using Scikit-learn \citep{pedregosa2011scikit}.
\\
\\ 
We use the term `DKL' to represent Longstaff-Schwartz algorithm with deep kernel learning (Algorithm \ref{alg:1}). The DKL model we adopted comprises a deep neural network structured as ($d$-1000-500-50-2), where $d$ denotes the number of underlying assets, and a sparse variational Gaussian process equipped with RBF kernel and 40 inducing points, unless stated otherwise. In some experiments, we evaluate both DKL40 and DKL200, where the numerals indicate the number of inducing points, to demonstrate the impact of the number of inducing points on pricing precision. 'LSM' represents the least squares Monte Carlo method that employs polynomial basis functions up to the second degree. Each experiment is conducted across 10 independent batches, each comprising $N=10,000$ simulated paths. The means and standard deviations of the results from these 10 batches are tabulated, with the numbers in parentheses indicating the standard deviations.

\subsection{Max options under GBM}
We first evaluate the performance of the proposed algorithm for $d$-dimensional max call options. Typically, pricing a max option is more challenging than a geometric basket option due to the highly non-linearity of the maximum function. 

\begin{equation}
    \text{payoff function:} \; h(S_{t})=\left( \max_{\nu=1,\dots,d}S_{t,\nu}-K\right)^{+}
\end{equation}
\begin{equation}
    \text{GBM:} \;  \frac{dS_{t,\nu}}{S_{t,\nu}}=(r-q)\, dt+\sigma \, dW_{t,\nu} \;\;\; \nu=1,\dots,d
\end{equation}
We adopt the same parameters as in reference \cite{refBecker2019}:
\\maturity $T=3$, strike price $K=100$, number of exercise dates $n=9$, risk-free interest rate $r=0.05$, equal dividend rate $q=0.1$, equal volatilities $\sigma=0.2$ and $W_t$ are $d$-dimensional Brownian motions with correlation $\rm{corr}(dW_{t,i},dW_{t,j})=\rho=0 \; \text{for}\; i\neq j$.

\subsubsection{Pricing Results for a Range of Dimensions and Discussion on the Estimated Continuation Value Surface}
\begin{figure}[htb]
  \centering
  \subfloat[$S_0=90$]
  {\includegraphics[width=0.5\textwidth]{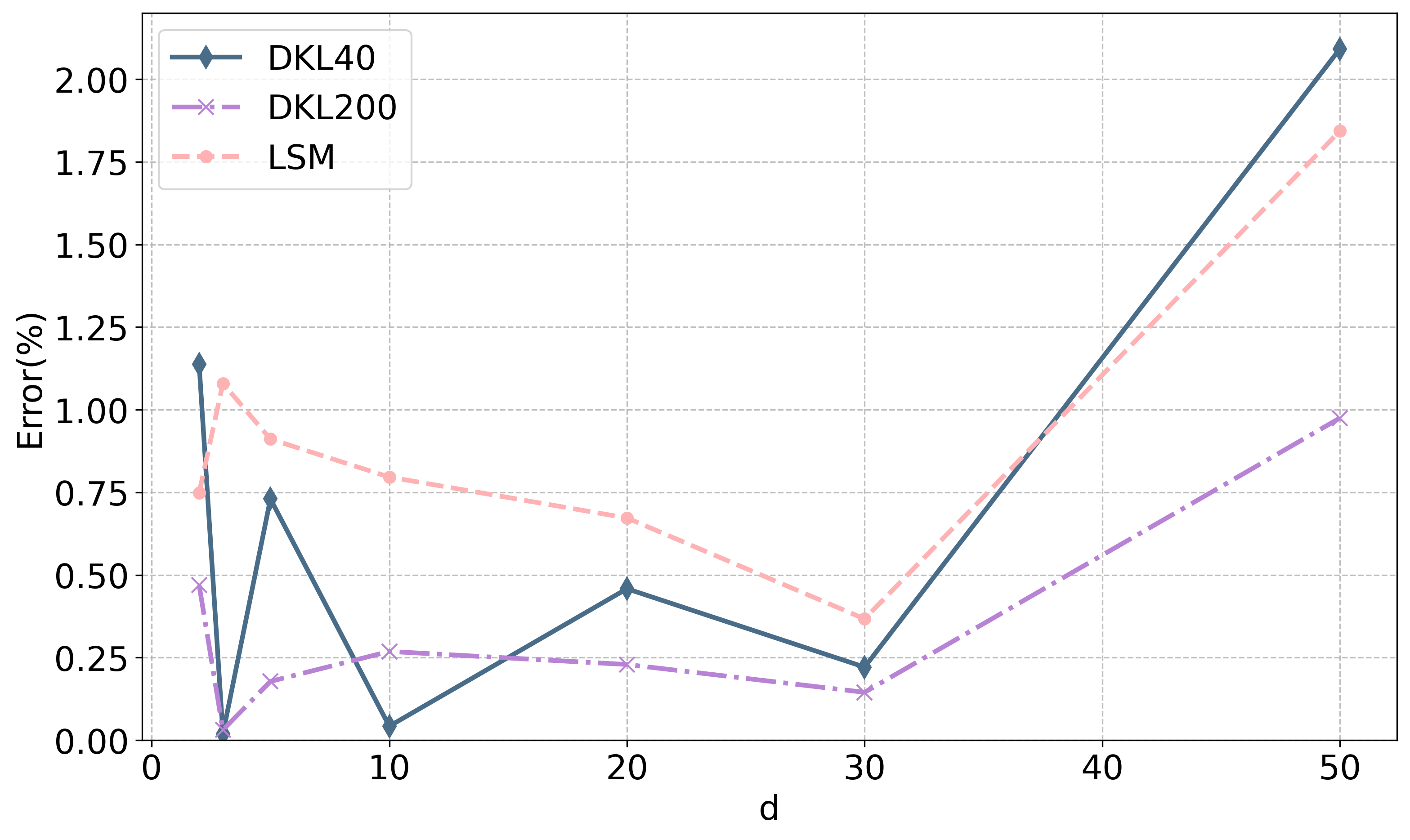}}
  \hfill
  \subfloat[$S_0=100$]
  {\includegraphics[width=0.5\textwidth]{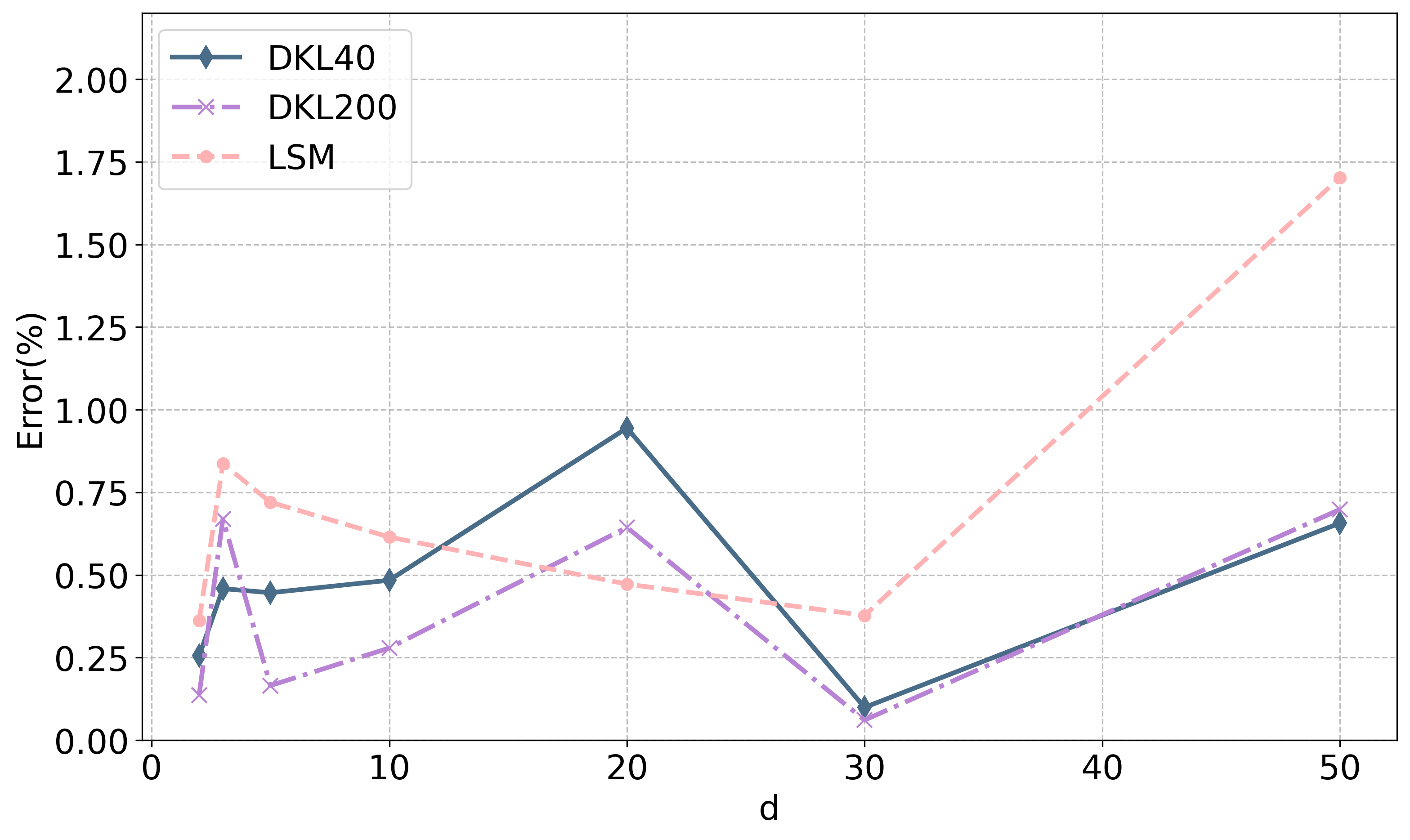}}
  \newline
  \subfloat[$S_0=110$]
  {\includegraphics[width=0.5\textwidth]{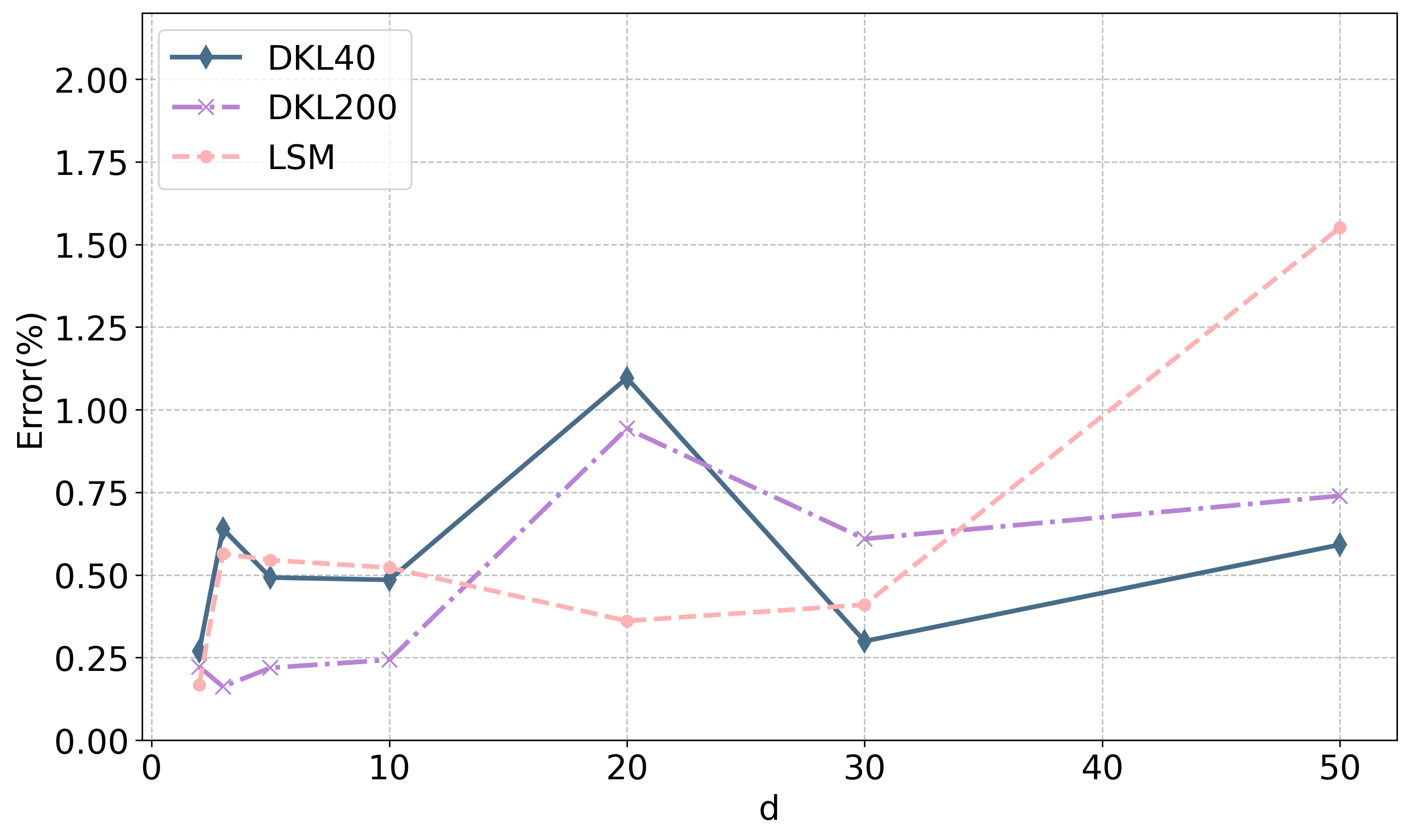}}
  \caption{Pricing errors under GBM for different dimensions}
  \label{fig_GBM}

\end{figure}

\begin{figure}[htb]
  \centering
  \subfloat[DKL]
  {\includegraphics[width=0.3\textwidth]{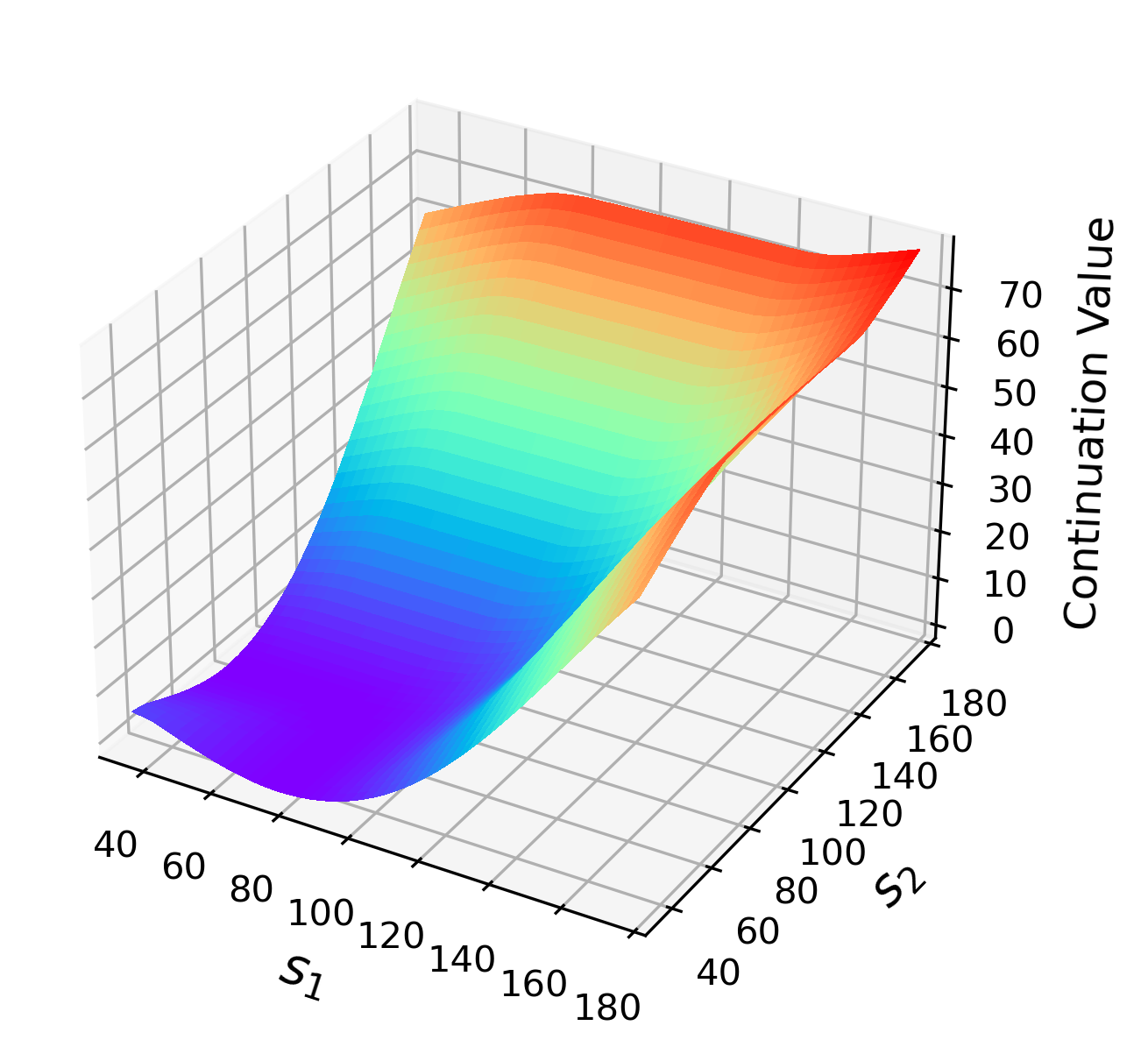}}\label{DKLA}
  \hfill
  \subfloat[LSM (up to the second degree)]
  {\includegraphics[width=0.3\textwidth]{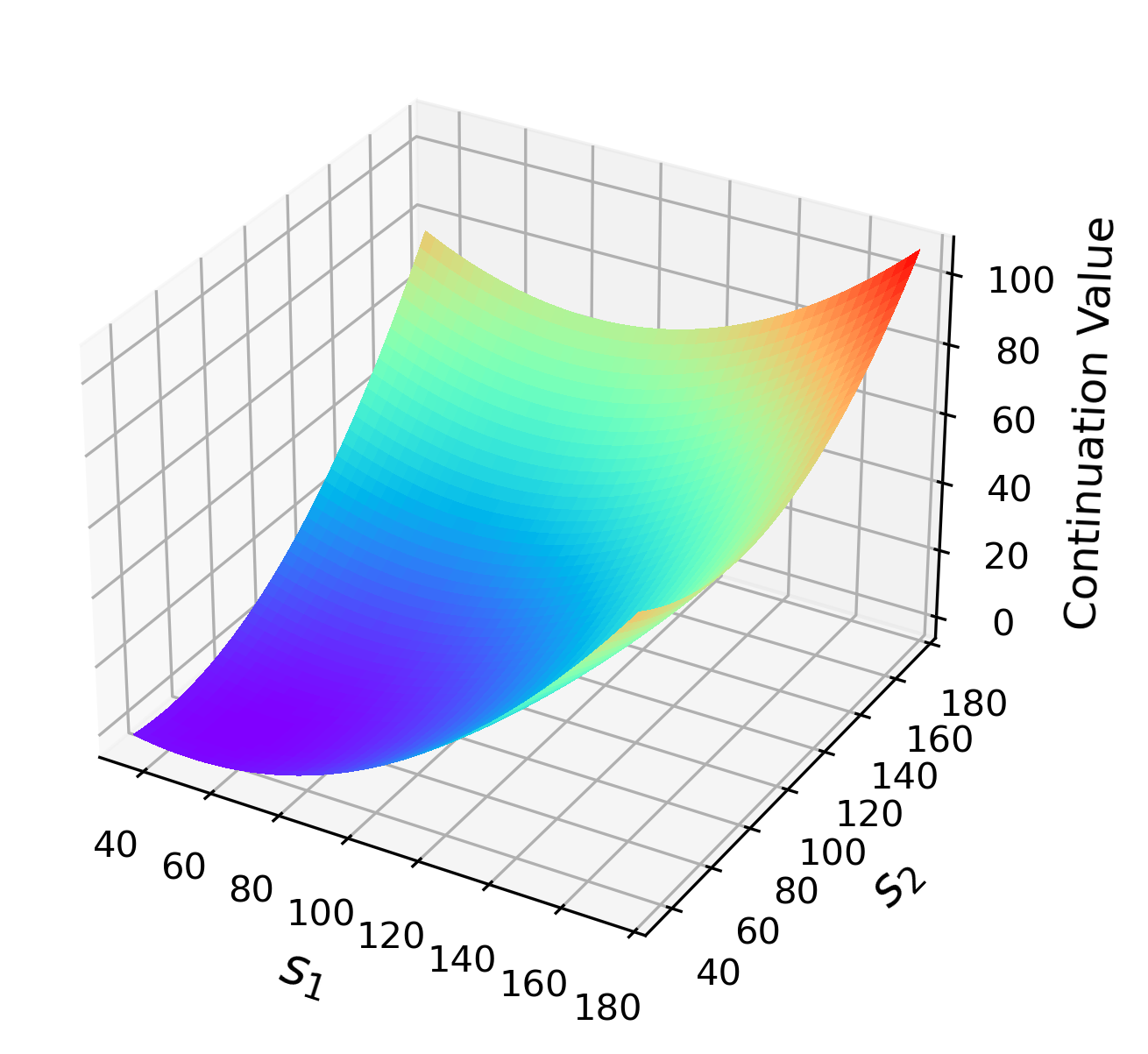}}
  \hfill
  \subfloat[LSM (up to the fifth degree)]
  {\includegraphics[width=0.3\textwidth]{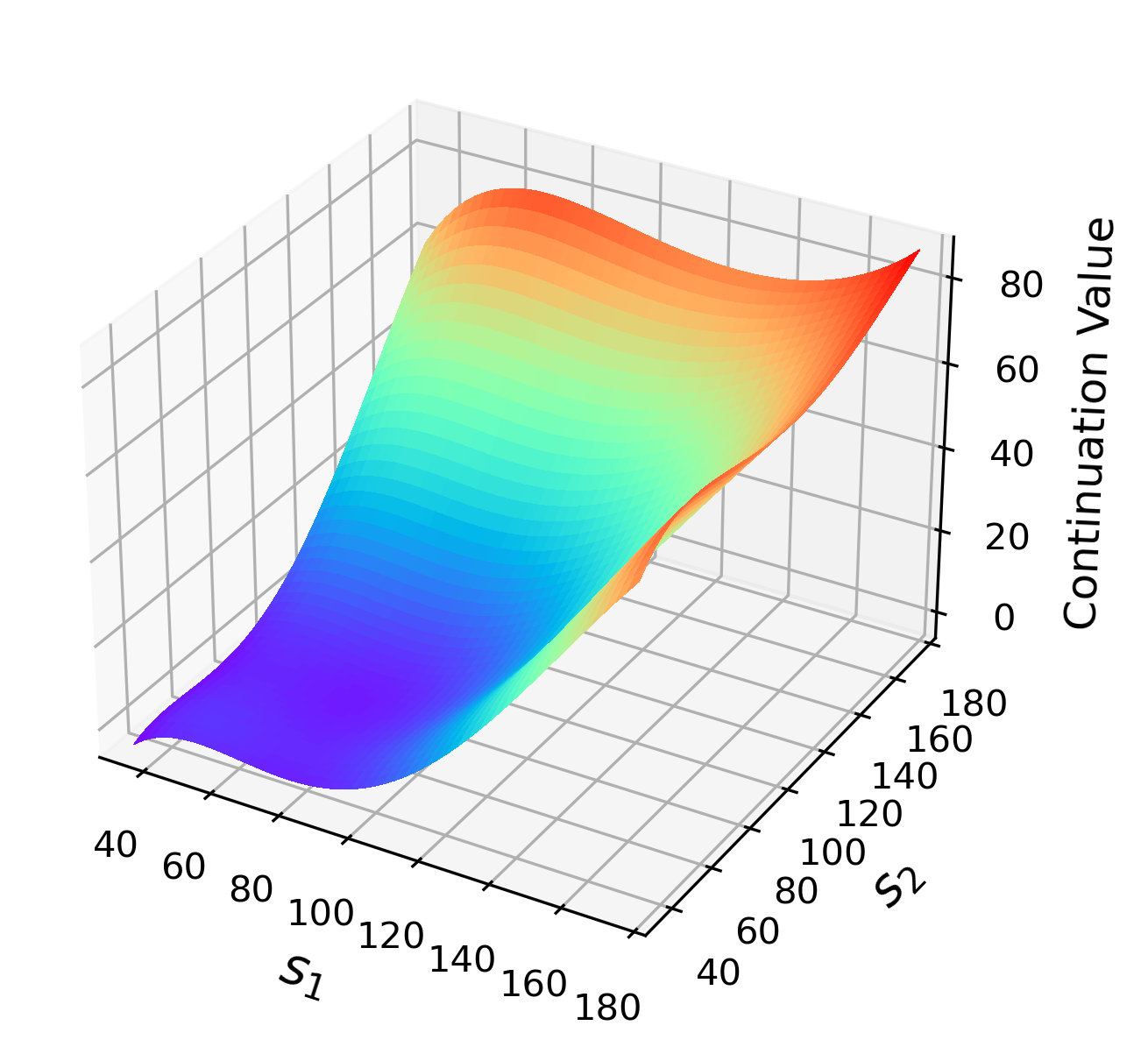}}
  \newline
\subfloat[DKL top-down view]
  {\includegraphics[width=0.3\textwidth]{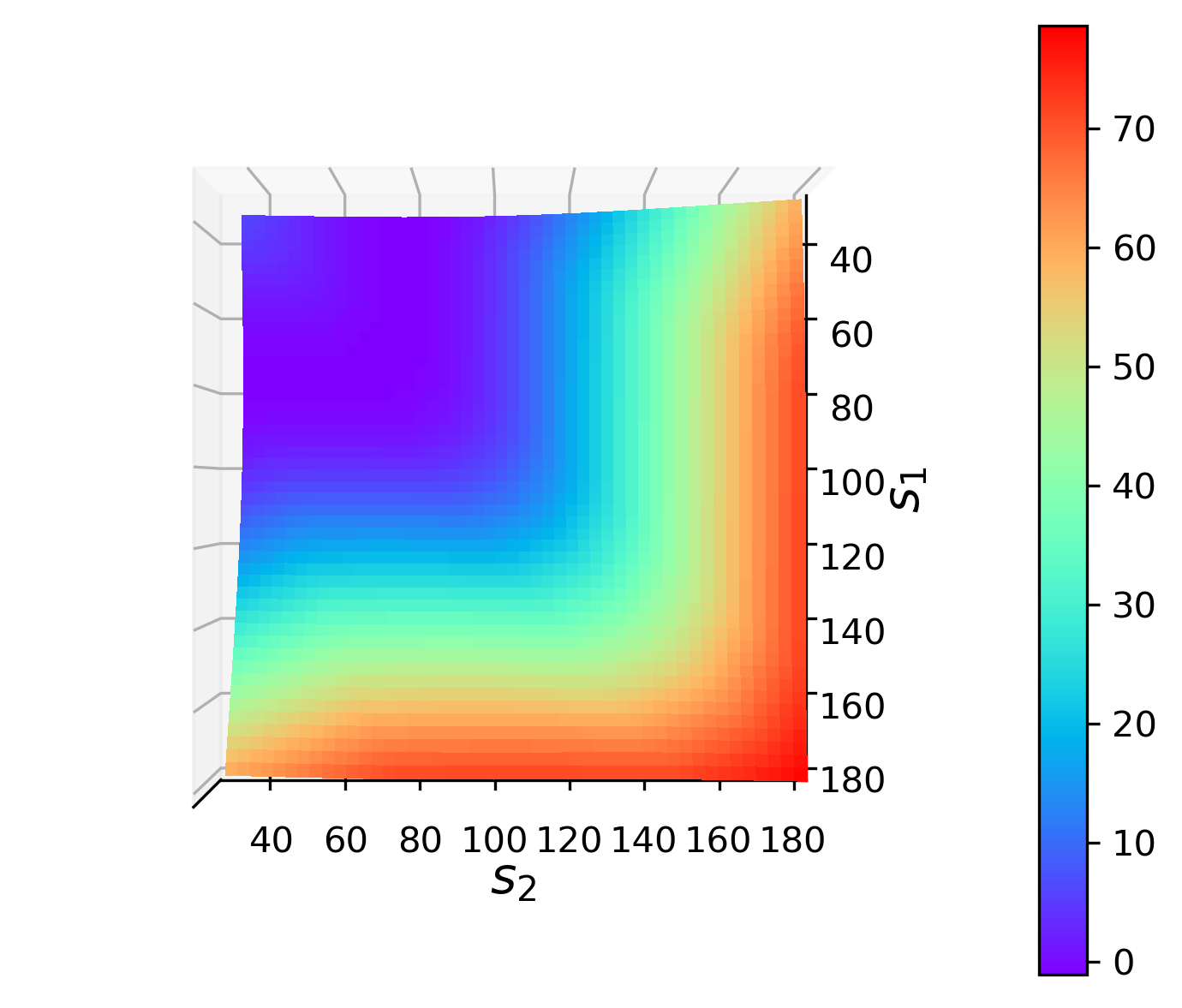}}
  \hfill
  \subfloat[LSM top-down view (up to the second degree)]
  {\includegraphics[width=0.3\textwidth]{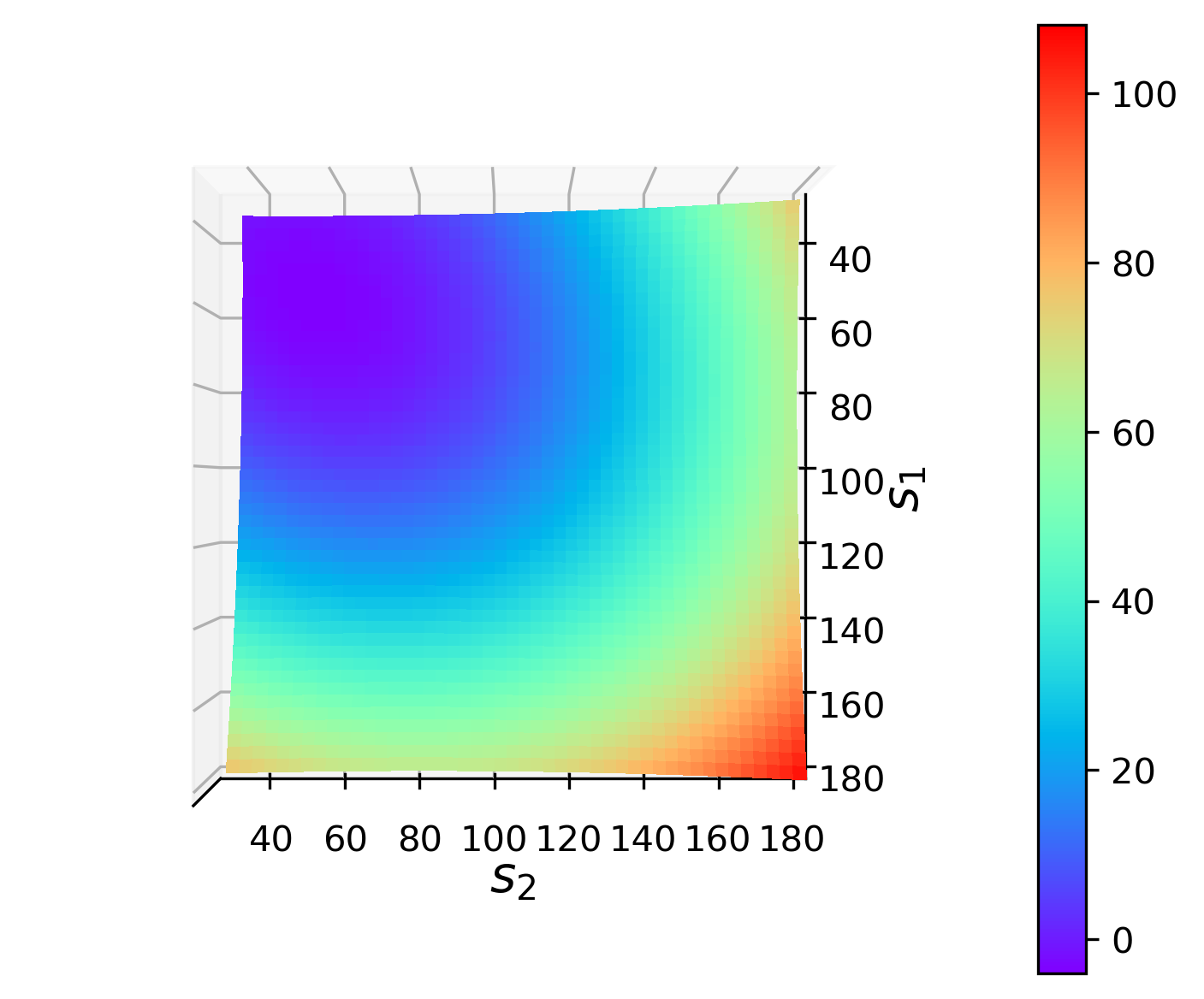}}
  \hfill
  \subfloat[LSM top-down view (up to the fifth degree)]
  {\includegraphics[width=0.3\textwidth]{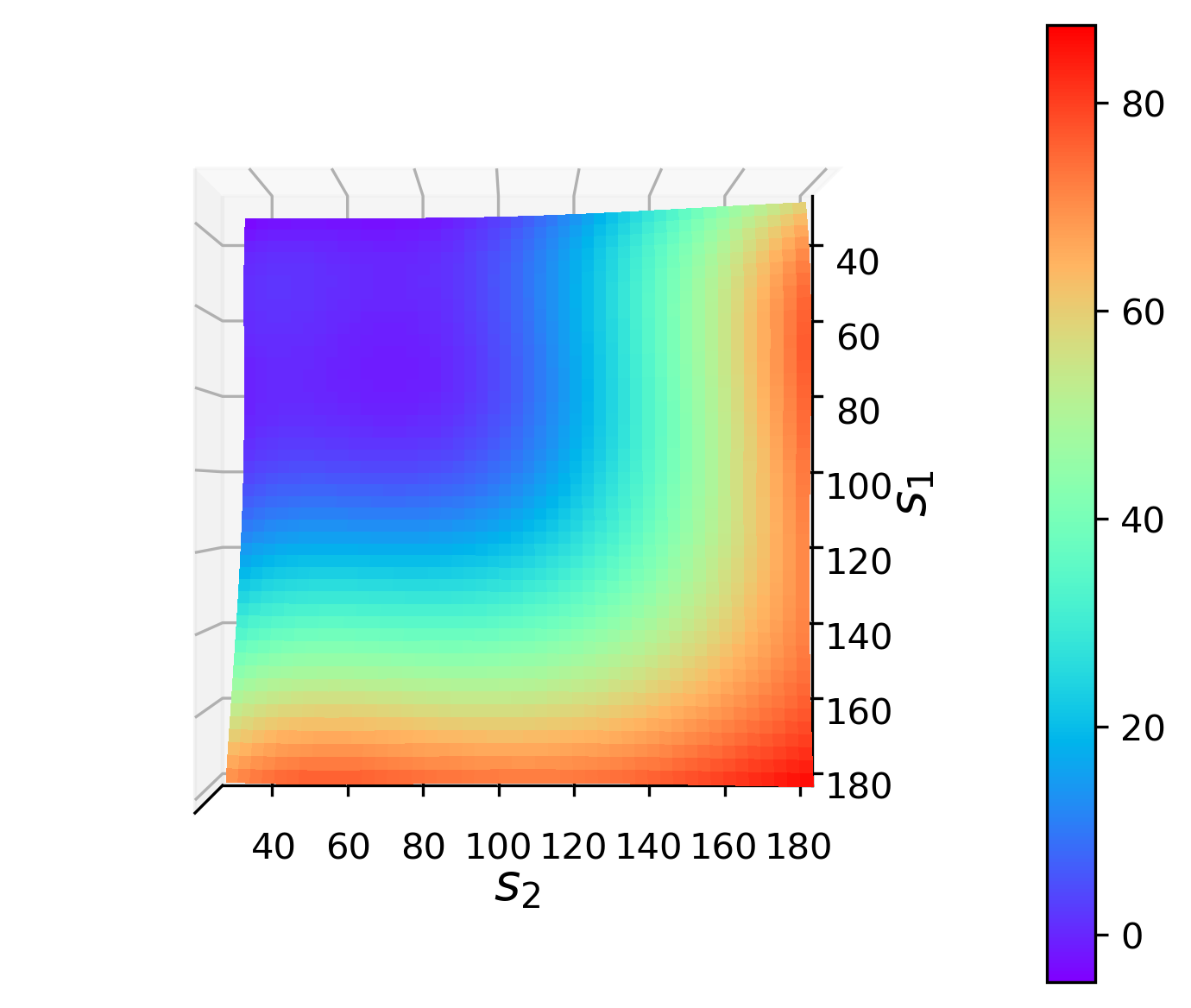}}

  \caption{Estimated continuation value surfaces of a two-dimensional max option under GBM at $t_{n-1}$, ($s_1,s_2\in [30,180]$ \;\;\;$S_0=100$)}
  \label{fig_continuation_value}
\end{figure}

Table \ref{table_GBM} and Figure \ref{fig_GBM} illustrate the performance of different methods for several initial price $S_0=90,100,110$ and up to 50 dimensions. Since there is no analytic solution for these option price, we adopted the the point estimation and $95\%$ confidence interval reported by \cite{refBecker2019} as our benchmarks, and these are denoted by ``BM point est." and ``BM $95\%$ CI" respectively. All errors are relative to the BM point est. It is evident that both DKL methods outperform the LSM in most scenarios; furthermore, the pricing errors of DKL200 are all lower that $1\%$.
\\
\\A possible reason for this phenomenon is that DKL methods may provide a more reliable estimated continuation value surface for the option during pricing, which we will illustrate with a numerical experiment. To visualize the continuation value surface, we consider only the 2-dimensional case in the following. The continuation value surface at $t_{n-1}$ is given by:
\begin{align}
    C(s_1,s_2)&=\mathbb{E}[V_n|S_{n-1,\,1}=s_1, S_{n-1,\,2}=s_2]\\
    &=\mathbb{E}\left[\max\left\{\max\left(S_{n,\,1}, S_{n,\,2}\right)-K,0\right\}|S_{n-1,\,1}=s_1, S_{n-1,\,2}=s_2\right]\label{eq_continuation}
\end{align}
From Equation \ref{eq_continuation}, we can deduce that the surface should have the following property: given fixed $s_2=c$ where $c$ is constant, the function $C_1(s_1)\coloneqq C(s_1,c)$ is monotonically increasing. Figure \ref{fig_continuation_value} illustrates the estimated continuation value surface and its corresponding top-down view at $t_{n-1}$, as obtained from various methods, including DKL, LSM with polynomial basis up to the second and fifth degrees. Observation of the figure reveals that the curve $C(s_1,180)$ of the estimated surface given by both LSM methods is not monotonic. Consequently, it is evident that the DKL method’s experimental outcomes are more consistent with the aforementioned property in comparison to the LSM methods.

\begin{table}[htb]
    \centering
    \resizebox{\linewidth}{!}{ 
    \begin{tabular}{ccccccc}
    \hline
    $d$  & $S_{0}$  & DKL40 & DKL200 & LSM  & BM point est.  & BM 95\% CI \\
    \hline
    2  & 90  &7.982 (0.227) &8.036 (0.111) &8.014 (0.118)             &8.074 & [8.060, 8.081]           \\
    2  & 100 &13.864 (0.173) &13.880 (0.160) &13.849 (0.181)         &13.899               & [13.880, 13.910]          \\
    2  & 110 &21.291 (0.205) &21.302 (0.171)        &21.313 (0.164)         &21.349               & [21.336, 21.354]           \\  &&&&&&
    \\
    3  & 90  &11.285 (0.154)&11.283 (0.166) &11.165 (0.155)             & 11.287              &[11.276, 11.290]           \\
    3  & 100 &18.604 (0.173)    &18.565 (0.178) &18.534 (0.181)               & 18.690              &[18.673, 18.699]           \\
    3  & 110 &27.397 (0.349)   &27.528 (0.216) &27.417 (0.168)             & 27.573              & [27.545, 27.591]           \\&&&&&&
    \\
    5  & 90  &16.766 (0.157)  &16.614 (0.331)&16.492 (0.166)         &16.644               &[16.633, 16.648]           \\
    5  & 100 &26.042 (0.287)    &26.116 (0.275) &25.971 (0.200)              &26.159               & [26.138, 26.174]           \\
    5  & 110 &36.591 (0.355)  &36.691 (0.244) &36.572 (0.228)           &36.772               & [36.745, 36.789]                \\&&&&&&
    \\
    10 & 90  &26.229 (0.421) &26.311 (0.398) &26.031 (0.146)            &26.240               &[26.189, 26.289]
           \\
    10 & 100 &38.151 (0.576)  &38.230 (0.288)    &38.101 (0.172)             &38.337               & [38.300, 38.367]               \\
    10 & 110 &50.639 (0.213)  &50.762 (0.405)      &50.620 (0.202)              &50.886               & [50.834, 50.937]
           \\&&&&&&
    \\
    20 & 90  &37.629 (0.334)  &37.715 (0.351)&37.548 (0.144)            &37.802               & [37.681, 37.942]\\
    20 & 100 &51.180 (0.842)  &51.335 (0.744)    &51.424 (0.136)            &51.668               & [51.549, 51.803]           \\
    20 & 110 &64.909 (0.295)  &65.009 (0.362)     &65.391 (0.138)             &65.628             & [65.470, 65.812]               \\&&&&&&
    \\
    30 & 90  &45.052 (0.549) &44.888 (0.459) &45.118 (0.226)        &44.953               &[44.777, 45.161]           \\
    30 & 100 &59.600 (0.636)   &59.623 (0.366)     &59.884 (0.228)        &59.659               & [59.476, 59.872]
           \\
    30 & 110 &74.145 (0.502)   &73.915 (0.532)         &74.673 (0.241)         &74.368               &[74.196, 74.566]
           \\&&&&&&
    \\
    50 & 90  &52.926 (0.658)  &53.530 (0.348) &55.053 (0.172)       &54.057               &[53.883, 54.266]           \\
    50 & 100 &69.278 (0.459)   &69.249 (0.686) &70.923 (0.207)          &69.736               &[69.560, 69.945]
           \\
    50 & 110 &84.957 (0.658)   &84.831 (0.537)      &86.789 (0.234)         &85.463               & [85.204, 85.763]           \\ \hline
    \end{tabular}}
    \caption{Pricing results under GBM for different dimensions}
    \label{table_GBM}
\end{table}

\subsubsection{The influence of hyperparameters}\label{subsubsection4.1.2}
\begin{figure}[htb]
    \centering
    \includegraphics[width=1\textwidth]{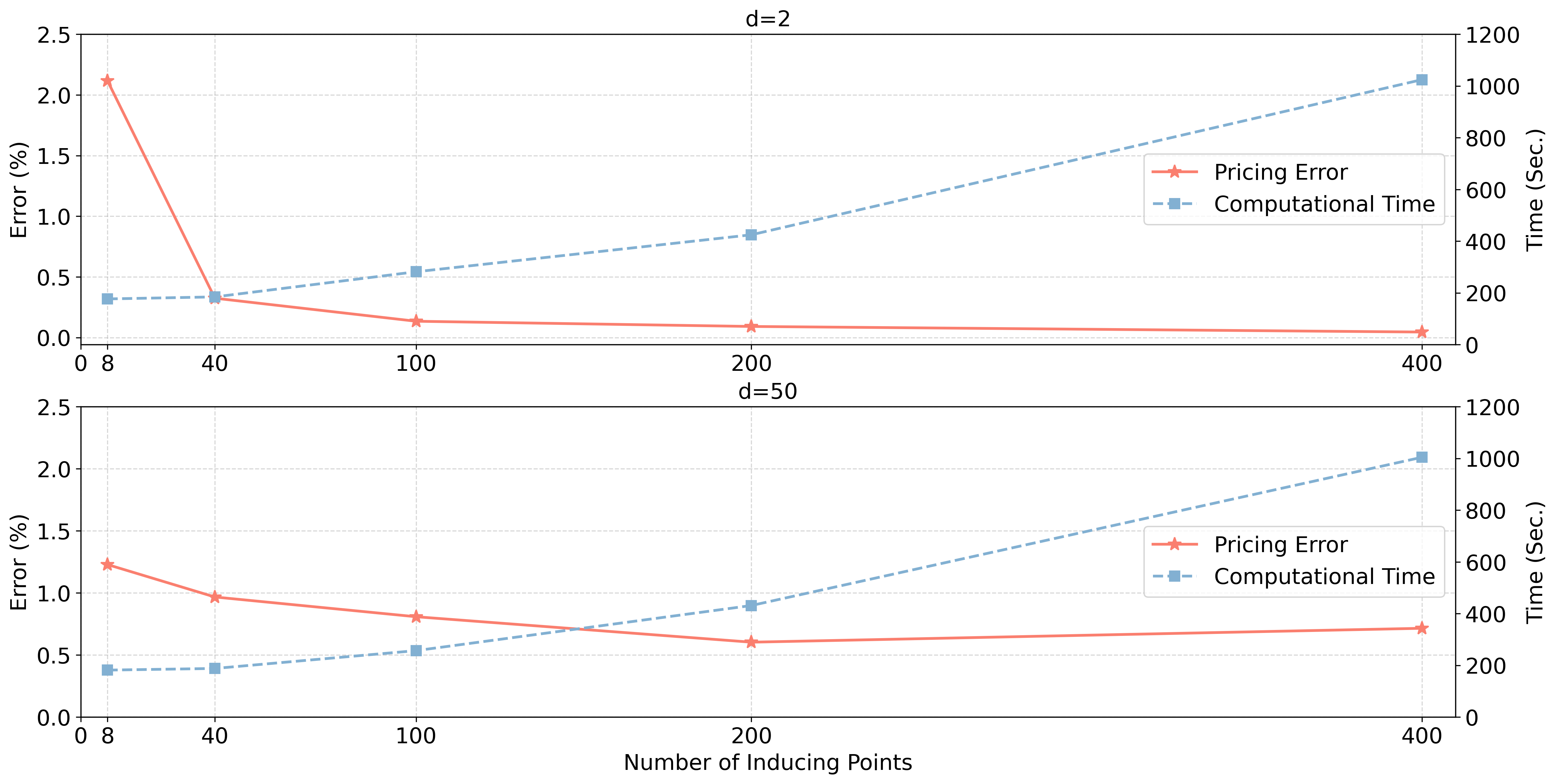}
    \caption{Pricing errors and computational time for DKL models employing varying numbers of inducing points ($S_{0}=100$)}
    \label{fig_inducing}
\end{figure}

\begin{figure}[htb]
    \centering
    \includegraphics[width=1\textwidth]{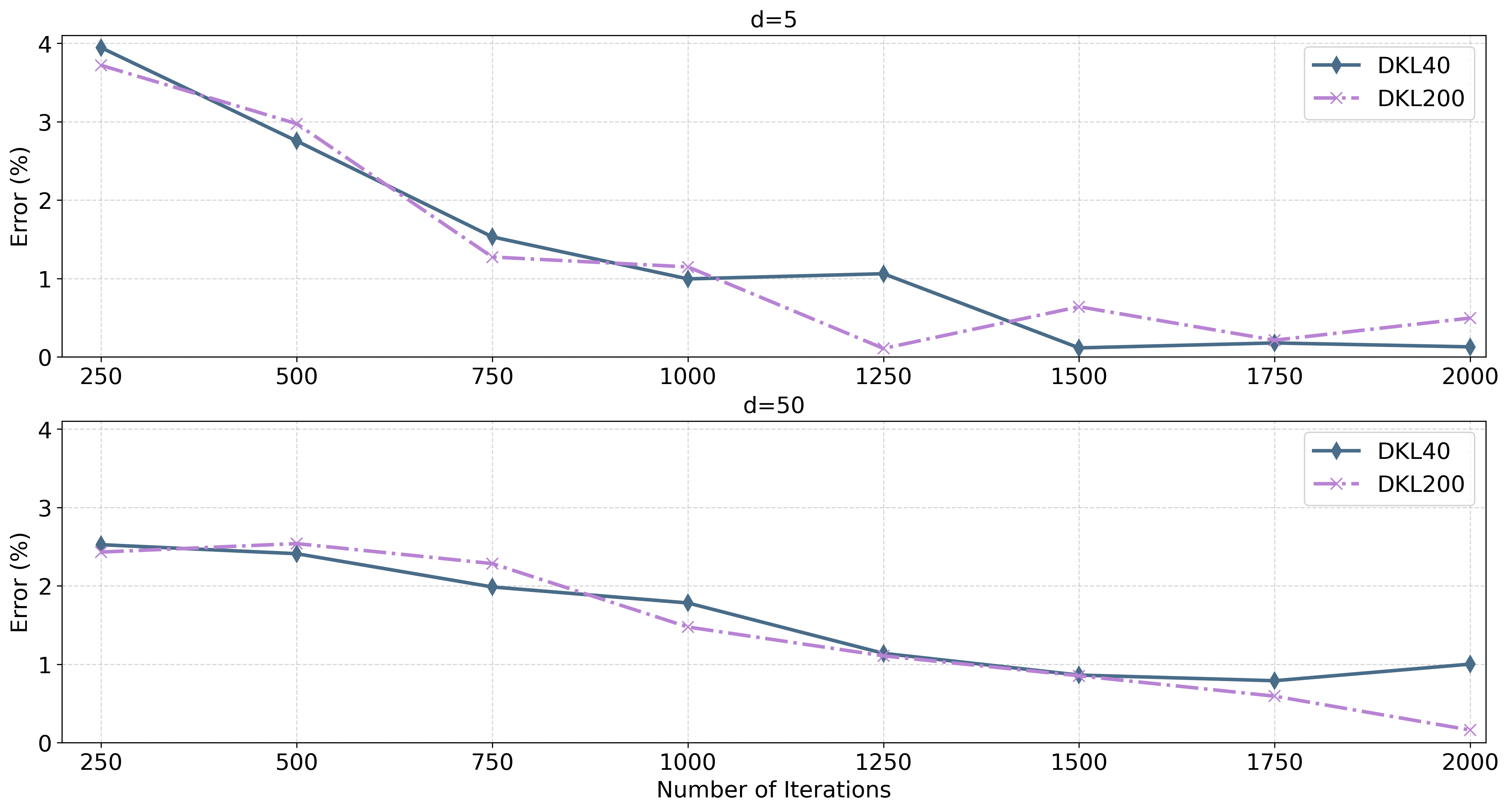}
    \caption{Pricing errors under GBM for different number of iterations ($S_{0}=100$)}
    \label{fig_iterations}
\end{figure}

Although the (hyper-)parameters in the Gaussian process part of the DKL model can be learned through a optimization problem, there are still several hyperparameters that must be determined initially, such as the number of inducing points, the width and depth of the feature extractor, and the number of iterations. This section presents various experiments to demonstrate their impact on pricing accuracy.
\\
\\
Figure \ref{fig_inducing} illustrates the pricing error and computational time of DKL methods with various numbers of inducing points in 2-dimensional and 50-dimensional cases.  It is noteworthy that there are no significant increases in computation time as the dimensions increase, leading to the conclusion that DKL models are not susceptible to the curse of dimensionality. Additionally, an increase in the number of inducing points may lead to higher computation time and improved accuracy. Table \ref{table_GPR} presents the results of the GPR with conventional RBF kernel. Together with Figure \ref{fig_inducing}, we notice that DKL models with a sufficient number of inducing points $(M \geq 40)$ are both faster and more accurate than GPR. Furthermore, the pricing errors of GPR exceed $(5\%)$ in the 50-dimensional case, suggesting the potential necessity of incorporating deep kernel learning into the Longstaff-Schwartz algorithm. Table \ref{table_feature} illustrates the influence of the feature extractor structure on pricing accuracy. According to Table \ref{table_feature}, the performance of the model with ($d$-31250-2) feature extractor is inferior to that of narrower feature extractors. This may be attributed to the heightened risk of overfitting in wider neural networks when estimating the continuation value, as a larger number of parameters requires training. The data also suggests that employing a neural network with a deep architecture as a feature extractor can significantly enhance accuracy. Figure \ref{fig_iterations} summarizes the behavior of the DKL models in both the 5-dimensional and 50-dimensional cases while varying the number of training iterations. It is observable that the pricing error remains relatively small in the $d=5$ case after 1250 iterations, whereas the $d=50$ case requires a greater number of iterations to converge.

\begin{table}[htb]
\begin{tabular}{|cc|cc|}
\hline
\multicolumn{2}{|c|}{$d=2$}          & \multicolumn{2}{c|}{$d=50$}         \\ \hline
\multicolumn{1}{|c|}{Error ($\%$)} & Time (Sec.) & \multicolumn{1}{c|}{Error ($\%$)} & Time (Sec.)\\ \hline
\multicolumn{1}{|c|}{0.780}      & 1125.3     & \multicolumn{1}{c|}{7.980}      &1102.0      \\ \hline
\end{tabular}
\caption{Pricing errors and computational time of Longstaff-Schwartz algorithm with GPR under GBM}
\label{table_GPR}
\end{table}

\begin{table}[htb]
\resizebox{\linewidth}{!}{
\begin{tabular}{lllllll}
\hline
                              &  & \multicolumn{2}{c}{$d=2$} &  & \multicolumn{2}{c}{$d=50$} \\
Structure of feature extractor       &  & Error ($\%$)       & Time (Sec.)     &  & Error ($\%$)      & Time (Sec.)       \\ \hline
$d$-50-2                        &  &  0.180           & 109.8          &  & 2.326            & 109.7           \\
$d$-250-2                       &  &  1.090           & 108.9          &  & 2.171            & 109.2           \\
$d$-1250-2                      &  &  1.077          &  132.5         &  &  1.928           &  134.4          \\
$d$-6250-2                      &  &  1.009           & 245.5         &  & 2.459            &  253.0          \\
$d$-31250-2                     &  &  1.571          &  741.1         &  &  2.607           &  860.9          \\
$d$-1000-500-50-2               &  &  0.290           & 173.9          &  & 0.519           &  175.7          \\
$d$-1000-500-250-125-50-2       &  &  0.156           & 211.1         &  & 0.708            &   204.4         \\
$d$-1000-500-250-125-50-25-10-2 &  &  0.453           & 205.1         &  & 0.761            &   219.2                 \\ \hline
\end{tabular}}
\caption{Pricing errors for DKLs with different feature extractor structures ($S_{0}=100$)}
\label{table_feature}
\end{table}

\subsubsection{Robustness Test}
\begin{figure}[htb]
    \centering
    \includegraphics[width=1\textwidth]{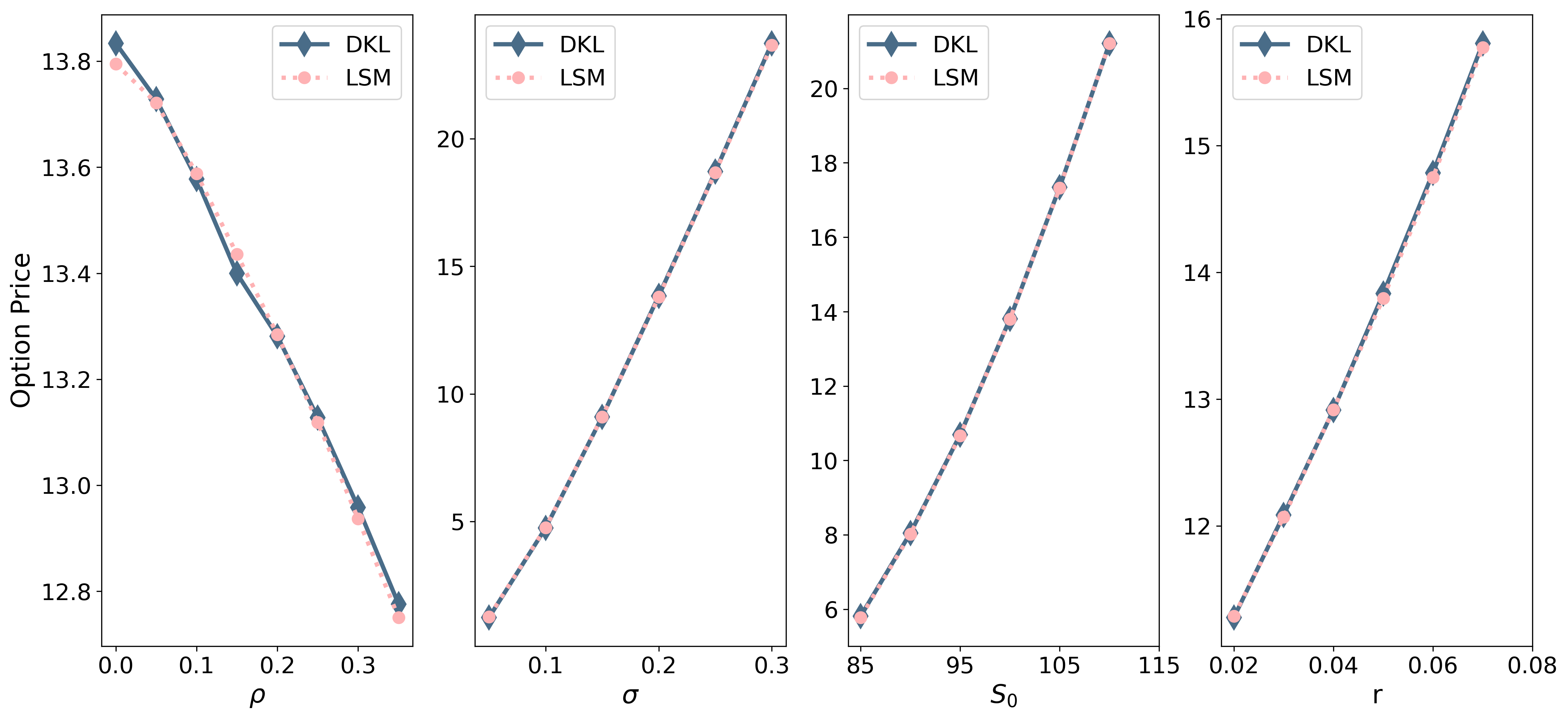}
    \caption{Pricing results under GBM for various parameters }
    \label{fig_robustness_test}
\end{figure}

In this section, we focus on the following question: once we have selected the hyperparameters mentioned in Section \ref{subsubsection4.1.2}, can the DKL method price options with different parameters accurately without re-selecting the hyperparameters? In the following experiments, the same DKL model has been used, which consists of a feature extractor structured as ($d$-1000-500-50-2), 40 inducing points and the number of iterations has been set to 1500.  We adopt the following default parameters: $d=2$, $S_0=100$, $T=3$, $K=100$, $n=9$, $r=0.05$, $q=0.1$, $\sigma=0.2$ and $\rho_{i,j}=\rho=0 \; \text{for}\; i\neq j$. In each experiment, one parameter from the set $\left\{\rho,\sigma,S_0,r \right\}$ changes, and the remaining parameters are set to their default values. As LSM is reliable in low-dimensional cases, we adopt its results as benchmarks. The results displayed in Figure \ref{fig_robustness_test} show that the DKL method provides accurate pricing in most scenarios and indicates that our method is robust to changes in parameters.

\FloatBarrier
\subsection{Geometric Basket Option under MJD}

\begin{figure}[htb]
  \centering
  \subfloat[Pricing Errors]
  {\includegraphics[width=0.5\textwidth]{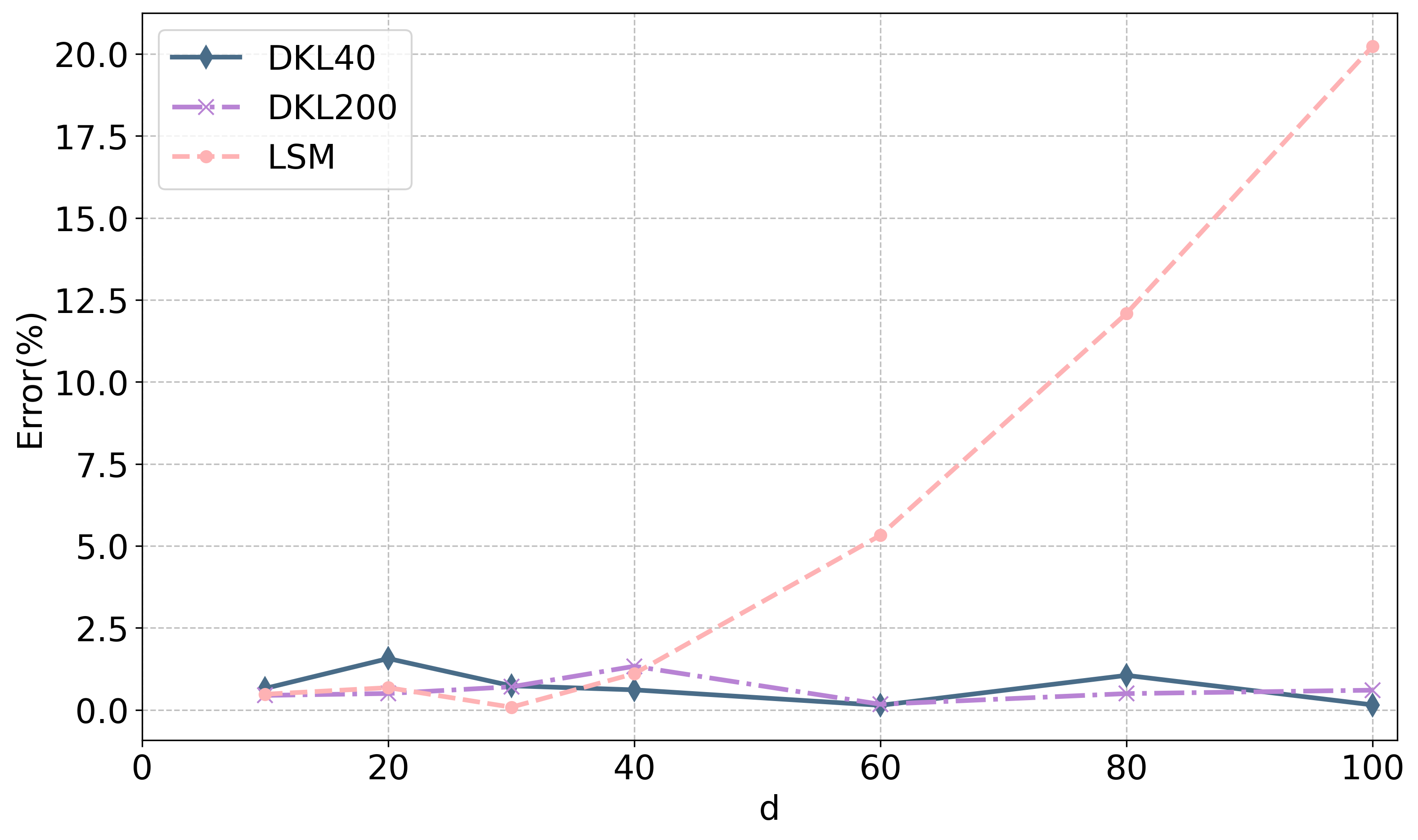}}
  \hfill
  \subfloat[Pricing Errors (zoomed in)]
  {\includegraphics[width=0.5\textwidth]{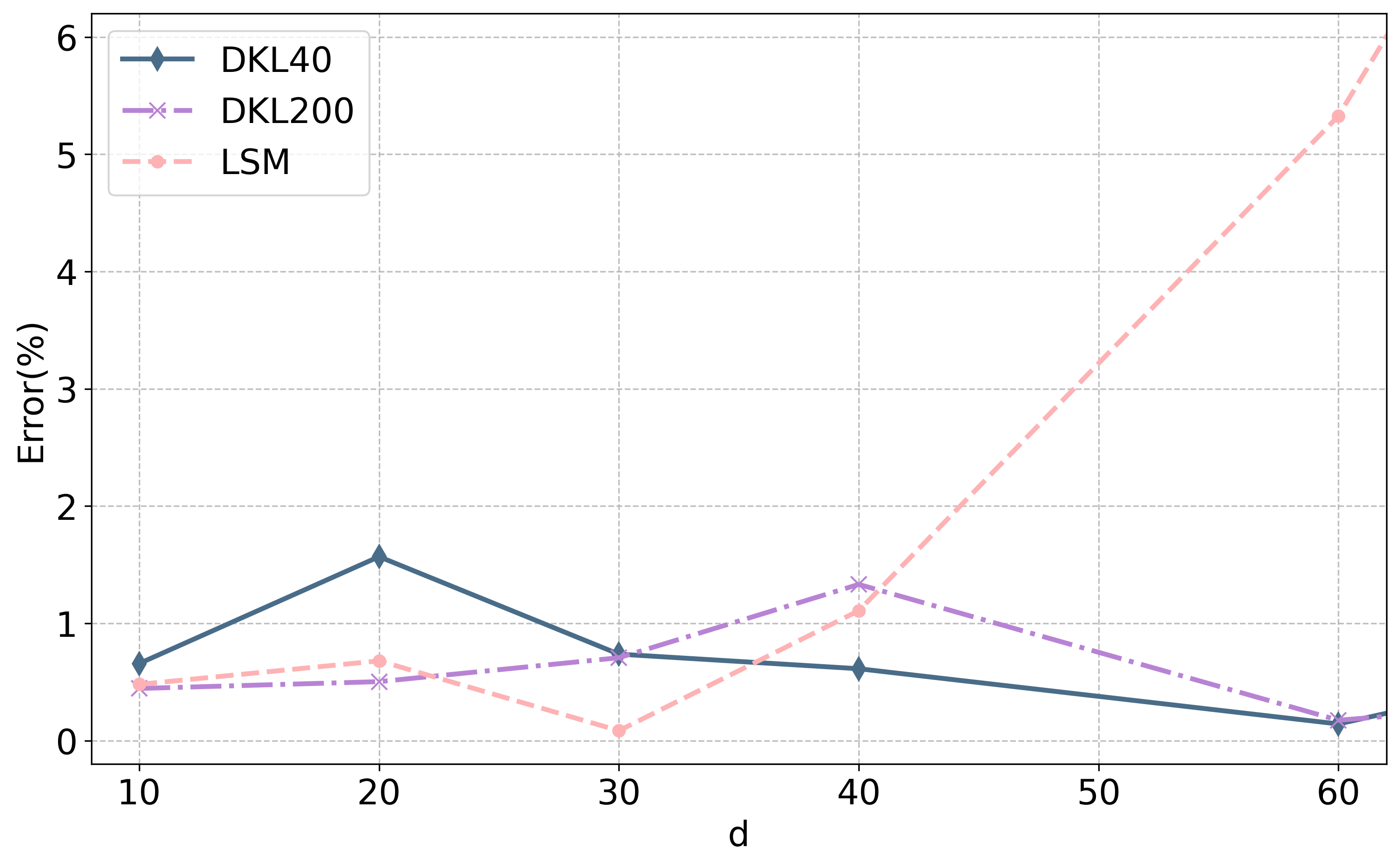}}

  \caption{Pricing errors under MJD for different dimensions}
  \label{fig_MJD}

\end{figure}

We consider a $d$-dimensional geometric basket put option where the underlying goods follow the MJD. We emphasize here that the hyper-parameters of the DKL model in this subsection are the same as those used in GBM case. The following example has been described in detail by \cite{refHu2020}:
\begin{equation}
    \text{payoff function:} \; h(S_{t})=\left( K-\prod^{d}_{\nu=1}S^{1/d}_{t,\nu}\right)^{+}
\end{equation}
\begin{equation}
    \text{MJD:} \;  \frac{dS_{t,\nu}}{S_{t,\nu}}=(r-q_{\nu}-\lambda^{J}\kappa_{\nu})\, dt+\sigma_{\nu} \, dW_{t,\nu}+\,dL_{t,\nu} \;\;\; \nu=1,\dots,d
\end{equation}
where the $W_{t,\nu}$ are standard Brownian motions with $\rm{corr}(dW_{t,i},dW_{t,j})=\rho_{i,j}$ and $L_{t,\nu}=\sum^{N(t)}_{n=1}(e^{Z^{J}_{\nu,n}}-1)$ is a compound Poisson Process with intensity $\lambda^{J}$ and $[Z^{J}_{1,n},\dots,Z^{J}_{d,n}]^{\intercal}$ following a multivariate Gaussian distribution $\mathcal{N}(\bm{\mu}^{J},\mathbf{\Sigma}^{J})$ with $\bm{\mu}^J=[\mu^{J}_{1},\cdots,\mu^{J}_{d}]^{\intercal}$ and $\mathbf{\Sigma}^{J}_{i,j}=\sigma^{J}_{i}\sigma^{J}_{j}\rho^{J}_{i,j}$. The expectation of the jump size is $\kappa_{\nu}=\exp(\mu^{J}_{\nu}+(\sigma^{J}_{\nu}/2))-1$.
\\
\\
We consider the same parameters that are adopted by \cite{refHu2020} under $d$ dimensions :
$S_{0,\nu}=K=40$, $T=1$, $n=10$, $\sigma_{\nu}=\sigma^{J}=1.5\cdot\sqrt{0.05}$, $r=0.08$, $\mu^{J}_{\nu}=-0.025$, $\lambda^{J}=5$, $q_{\nu}=\frac{1}{40}(1-1.5^{2})-5e^{-\frac{31}{32}}$, $\rho_{ij}=\rho^{J}_{ij}=\frac{\frac{4}{9}d-1}{d-1}$ for $i \neq j$. Notice that the correlation coefficients will change as the dimensions change. Under this settings, the benchmarks of the American options in all dimensions are equal to 6.995.
\\
\\
Table \ref{table_MJD} and Figure \ref{fig_MJD} display the pricing results up to 100 dimensions. It is clear that the LSM doesn't perform well as DKLs in high dimensional MJD. When $d\geq60$, the pricing errors of LSM are greater than $5\%$ and even reach $20\%$ in 100 dimensions. In contrast, the maximum error of DKL200 is lower than $1\%$ in most of the scenarios. This result shows that the DKL can be seen as a competitive alternative to LSM in pricing high dimensional American option under MJD.

\begin{table}[htb]
    \centering
    \begin{tabular}{cccc}
    \hline
    $d$   & DKL40 & DKL200 & LSM  \\
    \hline 
    10  &6.949 (0.132)     &6.964 (0.120)        &6.970 (0.051)               \\
    20  &6.885 (0.156)    &7.030 (0.107)          &6.981 (0.069)              \\
    30  &6.943 (0.140)    &7.044 (0.152)          &7.017 (0.066)              \\
    40  &6.952 (0.192)    &7.088 (0.141)          &7.140 (0.047)             \\
    60  &6.985 (0.219)   &7.007  (0.223)         &7.357 (0.083)            \\
    80  &7.069 (0.122)   &7.030  (0.264)         &7.814 (0.104)              \\
    100 &6.984 (0.205)    &7.037 (0.193)        &8.395 (0.050)            \\
    \hline
    \end{tabular}
    \caption{Pricing results under MJD for different dimensions (benchmark is 6.995)}
    \label{table_MJD}
\end{table}

\FloatBarrier
\section{Conclusion}
Valuing an American option involves an optimal stopping problem, typically addressed through backward dynamic programming. A key idea is the estimation of the continuation value of the option at each step. While least-squares regression is commonly employed for this purpose, it encounters challenges in high-dimensions, including a lack of an objective way to choose basis functions and high computational and storage costs due to the necessity of calculating the inverses of large matrices. These issues have prompted us to replace it with a deep kernel learning model. The numerical experiments show that the proposed approach outperforms the least-squares method in high-dimensional settings, and doesn't required specific selection of hyper-parameters in different scenarios. Additionally, it maintains a stable computational cost despite increasing dimensions. Therefore, this method holds promise as an effective solution for mitigating the curse of dimensionality.

\bibliography{reference}
\end{document}